\documentclass[apj]{emulateapj}

\usepackage{amssymb}
\usepackage{lscape}
\usepackage{longtable}

\def\lsim{\mathrel{\rlap{\lower4pt\hbox{\hskip1pt$\sim$}}
    \raise1pt\hbox{$<$}}}                
\def\gsim{\mathrel{\rlap{\lower4pt\hbox{\hskip1pt$\sim$}}
    \raise1pt\hbox{$>$}}}                

\shorttitle{}
\begin{document}

\shortauthors{Stark et al.}

\title{Keck Spectroscopy of $3<z<7$ Faint Lyman Break Galaxies: The Importance of Nebular
Emission in Understanding the Specific Star Formation Rate and Stellar Mass Density}


\author {Daniel P. Stark\altaffilmark{1,2}, Matthew A. 
  Schenker\altaffilmark{3},Richard Ellis\altaffilmark{3}, Brant Robertson\altaffilmark{1}, 
   Ross McLure\altaffilmark{4}, James Dunlop\altaffilmark{4}
}

\altaffiltext{1}{Department of Astronomy, Steward Observatory, University of Arizona, 
933 North Cherry Avenue, Rm N204, Tucson, AZ, 85721, dpstark@email.arizona.edu}
\altaffiltext{2}{Hubble Fellow}
\altaffiltext{3}{Cahill Center of Astronomy \& Astrophysics, California Institute of Technology,
MS 249-17, Pasadena, CA 91125}
\altaffiltext{4}{Institute for Astronomy, University of Edinburgh, Royal Observatory,
  Edinburgh, EH9 3HJ}

\begin{abstract} 

The physical properties inferred from the spectral energy distributions of $z>3$ galaxies have 
been influential in shaping our understanding of early galaxy formation and the role galaxies may
play in cosmic reionization.  Of particular importance is the stellar mass density at early times which
represents the integral of earlier star formation. An important puzzle arising from the measurements
so far reported is that the specific star formation rates (sSFR) evolve far less rapidly than expected 
in most theoretical models. Yet the observations underpinning these results remain very uncertain,
owing in part to the possible contamination of rest-optical broadband light from strong nebular
emission lines. To quantify the contribution of nebular emission to broad-band fluxes, we 
investigate the spectral energy distributions of 92 spectroscopically-confirmed galaxies 
in the redshift range $3.8<z<5.0$ chosen because the H$\alpha$ line lies within the {\it Spitzer}/IRAC 3.6$\mu$m filter.   
We demonstrate that the 3.6$\mu$m flux is systematically in excess of that expected from stellar continuum
alone, which we derive by fitting the SED with population synthesis models. No such excess is seen
in a control sample of spectroscopically-confirmed galaxies with $3.1<z<3.6$ in which there is no nebular
contamination in the IRAC filters. From the distribution of our 3.6$\mu$m flux excesses, we derive an H$\alpha$ equivalent width 
distribution and consider the implications both for the derived stellar masses and the sSFR evolution.  
The mean rest-frame H$\alpha$ equivalent width we infer at $3.8<z<5.0$ (270~\AA) indicates that nebular emission 
contributes at least 30\% of the 3.6$\mu$m flux and, by implication, nebular emission is likely to
have a much greater impact for galaxies with $z\simeq 6-7$ where both warm IRAC filters are contaminated.  
Via our empirically-derived  equivalent width distribution we correct the available stellar mass
densities and show that the sSFR evolves more rapidly at $z>4$ than previously thought, supporting 
up to a 5$\times$ increase between $z\simeq 2$ and 7.  Such a trend is much closer to
theoretical expectations.  Given our findings, we discuss the prospects for
verifying quantitatively the nebular emission line strengths prior to the launch of the James Webb Space Telescope.

\end{abstract} 

\section{Introduction}
\label{sec:intro}

Through detailed photometry of Lyman break galaxies (LBGs) undertaken with the 
Hubble and Spitzer Space Telescopes, much has been learned regarding the
physical properties of galaxies beyond a redshift $z\simeq$3.   Stellar masses and star formation rates have now 
been inferred from broadband photometric spectral energy distributions (SEDs) for thousands 
of galaxies spanning the redshift range $3<z<7$ (e.g., \citealt{Egami05,Eyles05,Labbe06,Eyles07,Stark07,Stark09,Ono10,
Gonzalez10,Labbe10a,Labbe10b,Gonzalez11a,Lee12,Reddy12,Curtis-Lake12}).  The stellar mass density derived from 
these studies has proven a useful integrated constraint 
on the contribution of galaxies to reionization (e.g., \citealt{Robertson10}), while the 
evolution of physical properties has provided insight into the processes which 
govern the assembly of early galaxies (e.g., \citealt{Finlator11,Dave11,Dave12}).

A potentially significant puzzle has recently emerged from these studies through 
measurement of the specific star formation rate (sSFR) at $z>2$.  Current observations 
demonstrate that between $z\simeq 2$ and $z\simeq 7$, the sSFR in galaxies of fixed stellar mass 
does not evolve strongly (e.g., \,\citealt{Stark09,Gonzalez10}), with recent estimates 
indicating at most a factor of two increase between $z\simeq 2$ and 7 (e.g.,
\citealt{Bouwens12b,Reddy12}).   This is in contrast to simple expectations from  
semi-analytic models and numerical simulations (e.g., \citealt{Weinmann11,Dave11,Dave12}) which 
predict that the sSFR should closely match the inflow rate of baryonic material. As this mass inflow rate 
is thought to increase 
with redshift as $\dot{M}/M\simeq (1+z)^{2.25}$ \citep{Neistein08,Dekel09}, we 
should expect nearly a 10$\times$ increase in sSFR in galaxies of fixed stellar mass over $2<z<7$, 
in marked contrast to the observations.

The physical cause of the discrepancy associated with the sSFR evolution remains unclear.    
As discussed previously (e.g., \citealt{Bouche10,Dutton10,Weinmann11,Dave11,Dave12,Reddy12}), 
a plateau in the redshift dependence of the sSFR would suggest star formation is more inefficient at 
$z>6$ than at $z\simeq 2$.  Various physical processes  might be invoked to impede star formation, 
such as the inefficient formation of molecular hydrogen in low metallicity galaxies (e.g., \citealt{Robertson08,Gnedin09,Krumholz12}) 
or an increase in the efficiency with which  cold gas is removed via large-scale outflows.   

Irrespective of any mechanism that might inhibit star formation at early times, it is difficult to reconcile such 
an inefficiency with the notion that early galaxies provide the ionizing photons responsible for reionization
\citep{Robertson10}. For example, the steep faint end slope of the ultraviolet luminosity function (UV LF) at 
$z>6$ \citep{Bouwens11} implies that star formation in low mass dark matter halos becomes {\it more efficient} 
at earlier times  (e.g., \citealt{Trenti10}), in contrast to the implications of the sSFR measurements.   

Given these difficulties, it is prudent that we reconsider the accuracy of the data that is used to infer the sSFR 
and its evolution.  The two basic ingredients are the star formation rates and the stellar masses. The $z>4$ 
measurements have indeed changed since the original articles (e.g., \citealt{Stark09,Gonzalez10}), 
mostly as a result of improved dust corrections following improved near-infrared photometry (e.g., \citealt{Bouwens12b}).  
The new dust corrections have served to increase the $z\simeq4$ sSFR measurements by a factor $\simeq$ 2.  
However, since negligible extinction is inferred at $z\simeq 6-7$, the sSFR still remains constant over
$4<z<7$ \citep{Bouwens12b} although a factor of 2 higher than at $z\simeq 2-3$ \citep{Reddy12}.

A potentially more important problem is the possible contribution of rest-frame optical nebular emission lines 
(e.g., [O II], [O III], H$\alpha$) to the broad-band fluxes used to infer the stellar masses. Such emission lines, 
could significantly affect the inferred amplitude of a Balmer Break, leading to an overestimate of the stellar mass
and thereby an underestimate of the sSFR. Figure 1 illustrates how the various nebular emission lines
contaminate the key photometric filters as a function of redshift.  Beyond $z\simeq4$, the key filters of
interest in the determination of stellar masses are the {\it Spitzer}/IRAC warm bands at 3.6$\mu$m and 4.5$\mu$m.
It is particularly striking that, at $z\gsim 5$, the strongest rest-frame optical nebular lines ([OIII]$\lambda5007$ 
and H$\alpha$) contaminate both {\it Spitzer}/IRAC filters. Although many $z\gsim 5$ galaxies are detected 
with {\it Spitzer} (e.g., \citealt{Egami05,Eyles05,Stark09,Labbe10a,Labbe10b,Gonzalez10,Gonzalez11a,Richard11b}),  
contamination by nebular emission could significantly affect the interpretation of their SEDs.

Accounting for nebular emission in the SEDs of high redshift galaxies has been considered
by several earlier works (e.g., \citealt{Schaerer09,Schaerer10,Ono10,deBarros12}).  In general
terms, their approach has been to use `forward modeling' techniques based on adding nebular emission 
contributions to stellar population synthesis models in order to demonstrate the possible implications of
its inclusion. However, such ``nebular+stellar" model fits cannot provide a precise unambiguous measure of nebular 
contamination for several reasons. Firstly, there are numerous uncertainties in how the contribution of 
nebular emission should be added. These include the nebular extinction law and ionizing photon 
escape fraction. Secondly, for galaxies at $z>5$, for which there is no uncontaminated 
measure of the stellar continuum (Figure 1), the uncertainties are particularly large. Finally, and perhaps
most importantly, without a spectroscopic redshift, addressing both the nebular contamination and
the photometric redshift of the galaxy from the same photometric data leads to great uncertainties; there
is no {\it a priori} indication of which photometric bands are contaminated by nebular emission.

Fortunately, by virtue of our deep Keck spectroscopic survey (Stark et al. 2010, Stark et al. 2011, Jones et al. 2012, 
Schenker et al. 2012) 
and our nebular+stellar population synthesis code (Robertson et al. 2010),
we can use the availability of HST-Spitzer SEDs to make progress in addressing this issue. While
the question of contamination by nebular emission at $z>5$ must await the infrared spectroscopic 
capabilities of James Webb Space Telescope, we can test our spectroscopic range $3.8<z<5.0$ for contamination
by H$\alpha$ in the {\it Spitzer}/IRAC 3.6$\mu$m broadband filter. Our approach follows that of
\cite{Shim11} who demonstrated that galaxies in this redshift window are typically significantly 
brighter at 3.6$\mu$m than at 4.5$\mu$m.  By comparing their flux density at 3.6$\mu$m to that 
expected from stellar continuum alone, \cite{Shim11} argued that many galaxies at $z>4$ 
show evidence for  strong H$\alpha$ emission, with typical EWs significantly greater than those 
seen at $z\simeq 2$.   

Here we seek to apply a similar technique to our spectroscopic sample 
\citep{Stark10,Stark11} with the goal of estimating the {\it distribution of H$\alpha$ equivalent widths} 
present in galaxies at $3.8<z<5.0$.  Equipped with this external constraint on the 
strength of nebular emission, we can then determine how stellar masses and the sSFR of $z>4$ 
galaxies are likely to be altered by emission line contamination.   In particular, we 
will explore whether our estimated degree of nebular contamination could be sufficient at the highest redshifts 
to permit a rapid rise in the redshift-dependent sSFR as expected from theoretical models.

The present paper is organized as follows.   In \S2, we discuss the selection
of the spectroscopic sample used in our analysis. 
In \S3, we introduce the details of our SED fitting procedure used to 
estimate the strength of nebular emission lines in the various
filters.  In \S4 we use our spectroscopic sample to estimate the 
equivalent width distribution of H$\alpha$ in the redshift range $3.8<z<5.0$ 
and then use these measurements to assess the impact of nebular 
emission on the derived stellar masses and star formation rates of 
$z>4$ galaxies.  In \S5, we discuss the impact that our findings have 
for the evolution in the integrated stellar mass density and specific 
star formation rates of galaxies at $z>4$.   

Throughout this paper we adopt a $\Lambda$-dominated, flat universe with
$\Omega_{\Lambda}=0.7$, $\Omega_{M}=0.3$ and $H_{0}=70\,h_{70}  {\rm
km\,s}^{-1}\,{\rm Mpc}^{-1}$. All magnitudes are quoted in the AB system \citep{Oke83}.

\begin{figure}
\epsscale{1.1}
\plotone{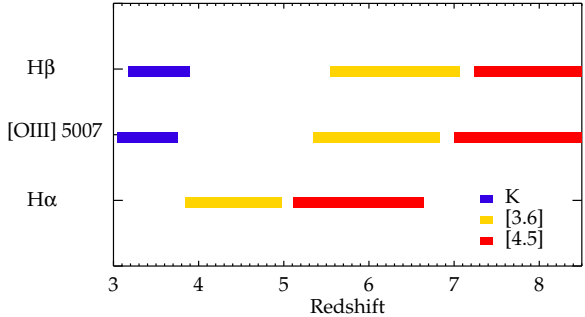}
\caption{Emission line contamination of broadband photometry.  Colored
stripes denote redshift ranges over which emission lines contaminate the 
K$_s$-band (dark blue), IRAC 3.6 $\mu$m (yellow), and IRAC 4.5 $\mu$m 
(red).    H$\alpha$ emission is expected in the 3.6$\mu$m filter at $3.8<z<5.0$.  
Note that at $5\lsim z\lsim 7$, both IRAC filters used for measuring stellar masses are contaminated by 
emission lines.  Beyond $z\simeq 7$, only the 4.5$\mu$m filter is contaminated 
by strong nebular emission.   }
\end{figure}

\section{Data}
\label{sec:sampleselection}

In this paper, we will focus our analysis on the interpretation of broad-band spectral 
energy distributions for a $z>3$ sample with known spectroscopic redshifts.  
The spectroscopic sample is drawn from earlier papers \citep{Stark10,Stark11}. 
Full details can be found in these articles, but we offer the reader a brief summary 
here.  Spectroscopy of $z>3$ LBGs in the two GOODS fields was undertaken at the 
Keck Observatory using the DEIMOS  
spectrograph \citep{Faber03}.  As discussed in \cite{Stark10}, LBGs were 
selected using standard `dropout' criteria (e.g., \citealt{Bouwens07,Stark09}) to a 
limiting magnitude of $z_{850}\simeq 27$ 
using the GOODS v2 public photometric catalogs (e.g., \citealt{Giavalisco04}).  
Taking advantage of a similar spectroscopic campaign undertaken using the FORS2 in GOODS-South 
(e.g., \citealt{Vanzella09}), we retrospectively constructed a VLT sample
using the same photometric criteria.  The combined Keck plus VLT survey comprises 157 galaxies 
in the redshift range $3.8<z<5.0$ which satisfy the dropout criteria. 
As we will discuss below, only a subset of these will be used in our analysis.

A key requirement for the derivation of stellar masses and specific star formation rate is
precise broad band photometry from which SEDs for galaxies of known spectroscopic redshifts
can be determined.  In GOODS-South, we use the public release of the Wide Field Camera 3 (WFC3) imaging from 
the CANDELS Multi-Cycle Treasury Program \citep{Grogin11,Koekemoer11} and our 
own reduction (see \citealt{McLure11}) of the Early Release Science 
(ERS) campaign (e.g., Windhorst et al. 2011).  Colors were computed with respect to the z$_{850}$ 
flux using matched apertures with up-to-date zero points, and total WFC3 magnitudes 
were derived by combining the measured colours with 
the total z$_{850}$-band flux.  K$_s$-band photometry is taken from deep ISAAC imaging 
\citep{Retzlaff10} following the procedure discussed in \cite{Stark09}.   For GOODS-N, we 
use near-infrared imaging obtained from CFHT/WIRCAM \citep{Wang10}.

The rest-frame optical at $z>4$ is probed by the deep {\it Spitzer}/IRAC (Fazio et al. 2004) 
imaging of GOODS-S  and GOODS-N (Dickinson et al. 2012, in prep).   In particular, 
the 3.6$\mu$m (hereafter [3.6]) and 4.5$\mu$m (hereafter [4.5]) are the most useful, as 
the longer wavelength filters are typically not sensitive enough to detect most $z>4$ galaxies.
As in Stark et al (2009), we focus primarily on the subset of ACS-selected galaxies whose IRAC 
fluxes are not contaminated significantly by neighboring sources. The IRAC magnitudes are 
measured in apertures 2.4 arcsec 
in diameter and to account for flux falling outside this aperture, we apply a 0.7 mag aperture 
correction derived from a sample of isolated point sources.  Recognizing that
selecting only isolated IRAC sources limits the size of our eventual sample, 
we included IRAC flux measurements for galaxies in GOODS-South 
from the MUSIC catalog \citep{Grazian06,Santini09}.  These fluxes 
rely on a deconfusion  procedure to extract fluxes from sources with contaminating 
neighbors.   A comparison between the 
two photometry methods reveals consistency for our isolated sample, with a standard 
deviation of 0.19 mags and no systematic offset.

In total, we have 92 galaxies in the range $3.8<z<5.0$ with measured IRAC photometry.  
In our analysis, we will focus on the subset of 45 galaxies with confident ($>$5$\sigma$) 
4.5$\mu$m detections (see Table 1), as without an accurate measure of the 4.5$\mu$m flux it is 
impossible to infer the expected stellar continuum from population synthesis models. 
The objects with deconfused GOODS MUSIC photometry make up 60\% of the final 
sample.  

Some caution must be exercised when applying inferences from a spectroscopic sample to the 
parent photometric population.    In attempting to infer the typical level of rest-optical nebular 
contamination, we must be particularly careful that we do not bias our sample toward strong 
Ly$\alpha$ emitting galaxies, a population which might have larger than average sSFR and 
H$\alpha$ EW.    For faint galaxies ($\rm{z_{850}>25.5}$), the spectroscopic sample of Stark et al. (2010) is 
indeed biased toward Ly$\alpha$ emitters.    But by requiring a 5$\sigma$ detection in the [4.5] band, we limit our sample to 
brighter systems (average $z_{850}$-band magnitude of 25.0) for which we are more complete spectroscopically.  
Indeed the percentage of galaxies for which we measure Ly$\alpha$ in emission is actually only 46\%, highlighting the 
fact that many galaxies in this bright subset are instead confirmed via the combination of Ly$\alpha$ in absorption and 
metal absorption lines (e.g., Jones et al. 2012).     Furthermore, the fraction of galaxies in this subset with 
strong ($\rm{EW>50}$~\AA) Ly$\alpha$ emission (5\%) is similar to that 
measured for galaxies in this M$_{\rm{UV}}$ and redshift range (6\% in Stark et al. 2010) adding confidence that the 
sample we use in this paper is not likely to be strongly biased toward nebular emitters and appears fairly representative 
of the photometric population.   

\section{Population Synthesis Modeling}

\subsection{Modeling Procedure}

Our goal is to quantify the nebular contribution through analysis of the SEDs of 
a large spectroscopic sample at $3.8<z<5.0$.   The advantage of our technique
is that, for these sources, we can predict the exact wavelengths of rest-frame 
optical emission lines and thereby remove ambiguities associated with determining
the photometric redshift simultaneously from contaminated broadband photometry. 

Previous attempts to assess the impact of nebular emission on broadband photometry 
have utilized models which include the contributions from both nebular and 
stellar emission.   For the reasons outlined in \S1, our analysis will focus instead on 
models containing only stellar continuum.     The stellar continuum predictions are based on the 
models of Bruzual \& Charlot  (2003) and the 
technique we will adopt is mostly similar to that described in detail in Stark et al  
(2009).     However, we also investigate how including 
nebular emission affects the derived physical properties.  To do so, we make 
use of the code described in \cite{Robertson10} to which the
interested reader is referred.    In this code, line emission is calculated from the 
number of ionizing photons per second, which is provided 
as output from our population synthesis models.

\begin{figure*}
\epsscale{0.9}
\plotone{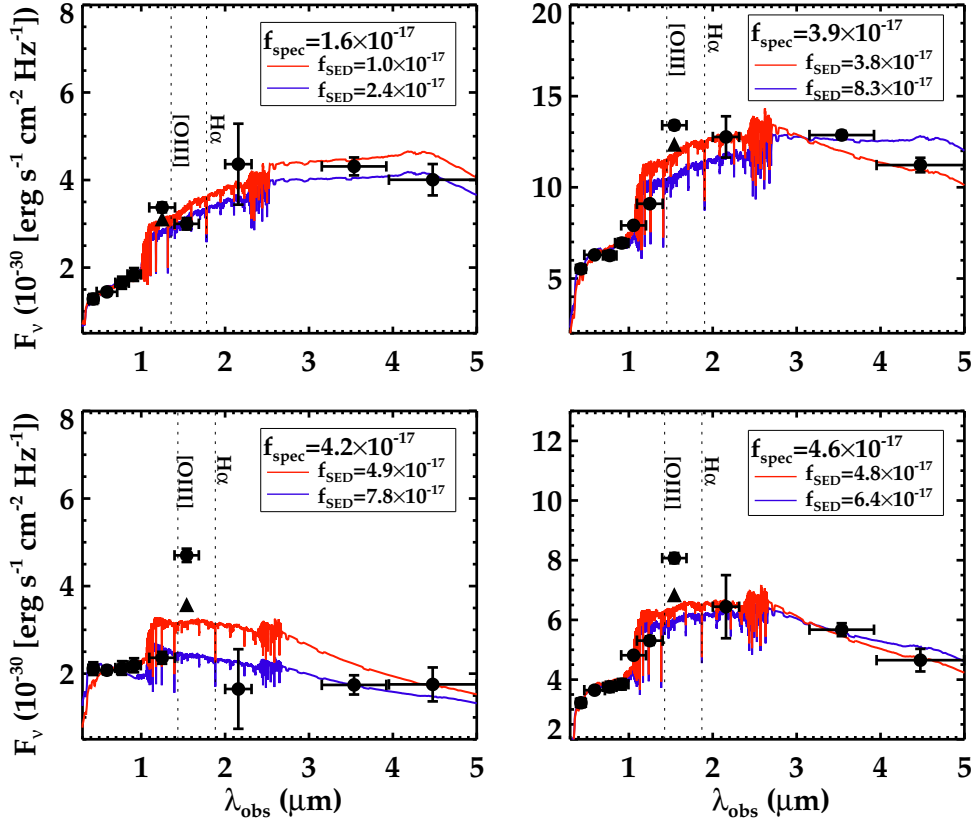}
\caption{Verification of the flux excess method of measuring emission lines strengths. The four panels 
show SEDs of spectroscopically-confirmed galaxies with $1.8<z<2.3$ for which measurements of 
emission line fluxes are available from the WFC3 grism study of \cite{Trump11}.  
The flux in the broadband filter contaminated by [OIII] is greater than that expected from 
the best-fitting stellar continuum models in each of the four SEDs.   The two displayed stellar 
continuum models correspond to fits excluding the contaminated filter (blue bottom curve) and 
fits including all available optical and near-IR filters (red top curve).   The flux density obtained 
by subtracting the directly-measured emission line contribution to the contaminated broadband 
flux is shown with a  triangle.    In each panel, we provide the [OIII] emission line flux measured 
with the WFC3 grism (f$_{\rm{spec}}$) and the the emission line flux inferred from the photometric 
excess of the contaminated filter with respect to the stellar continuum models (f$_{\rm{SED}}$) in units 
of erg/cm$^2$/s.  
 }
\end{figure*}

In the Robertson et al. (2010) code, the intensities of hydrogen lines are computed from the values tabulated in 
Osterbrock \& Ferland (2006), assuming case B recombination.  We compute the intensities
of the lines of common metallic species from the empirical results of Anders et al. (2003), 
assuming a gas phase metallicity of $Z = 0.2~ Z_{\odot}$, similar to that measured for galaxies 
at these redshifts (Maiolino et al. 2008, Jones et al. 2012).     The recombination line 
luminosities are calculated assuming that the ionising photon escape fraction, f$_{\rm{esc}}$, is 0.2 (see Shapley 
et al. 2011 for a discussion of expected escape fractions). 
Since we do not rely on the nebular models for our primary conclusions, this assumption does 
not affect our results.  The continuum contribution from bound-free, free-free, and two photon continuum emission is 
also calculated following Osterbrock \& Ferland (2006).  The full nebular template is then added to each 
stellar continuum model, which is then used to calculate the synthetic fluxes used in 
our SED fitting code.

Since our sample has the virtue of precise spectroscopic redshifts, we do not fit the photometric bands spanning 
the Ly$\alpha$ forest and Ly$\alpha$ emission lines, both of which vary significantly 
from source to source at any given redshift.   For the redshift range which we are primarily 
interested in ($3.8<z<5.0$), this leaves 7-8 (largely) independent photometric constraints on the SED in 
GOODS-S and 6 constraints on GOODS-N.   For consistency with the earlier literature, we 
consider a Salpeter (1955) initial mass function with 0.1-100 M$_\odot$.  Given the 
relatively small number of constraints on the SED, we utilize a moderately restricted 
grid, varying only the age, dust reddening, and normalization factor.   We fix the star formation 
history as either constant or rising with time following the t$^{1.7}$ power law inferred in Papovich et al. (2011).   
This restricted grid of SFH is supported by 
the results of Reddy et al. (2012) that demonstrate that at $z\simeq 2$, the star formation rates inferred from 
exponentially-declining star formation history models do not agree with those measured from 
the observed IR and UV fluxes.   Nevertheless we have verified that our results would not be affected if we had adopted 
exponential decay models.  Finally we utilize sub-solar metallicity (Z=0.2~Z$_\odot$) motivated by the observations 
discussed  above.  

\begin{figure}
\epsscale{1.2}
\plotone{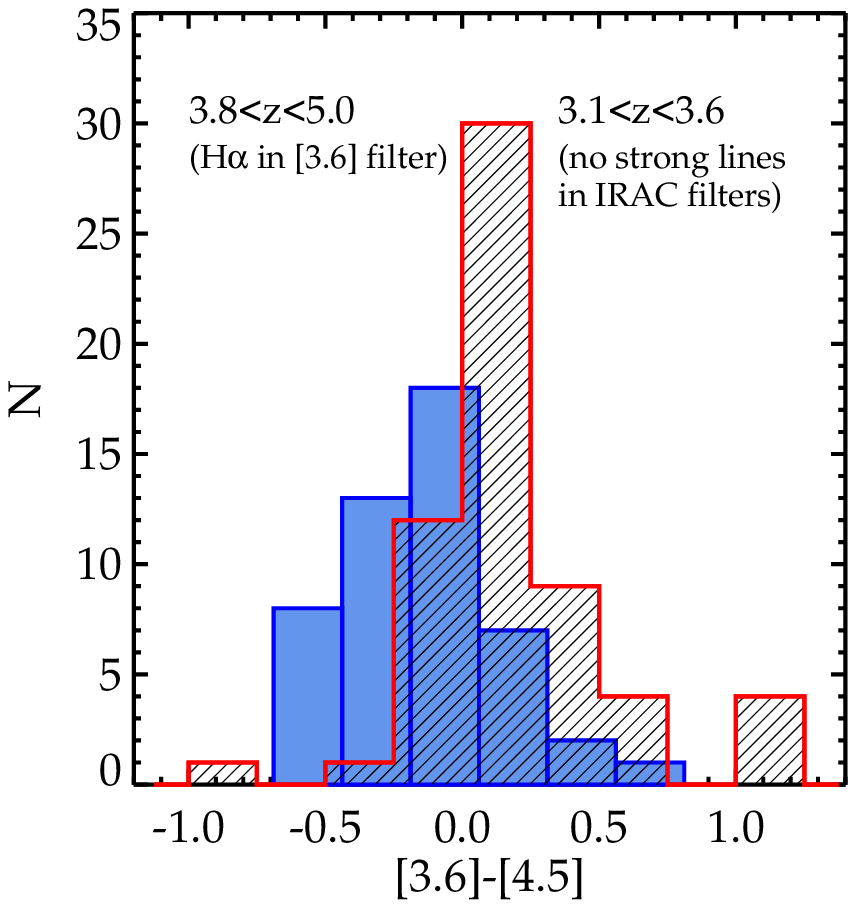}
\caption{The effect of emission lines on broadband colors. The distribution of [3.6]-[4.5] colours for 
galaxies at $3.8<z<5.0$ (blue filled histogram) and at $3.1<z<3.6$ (red shaded histogram). 
The colours in the lower redshift sample reflect the reddened stellar continuum, as both 
filters are free from strong emission lines.  In contrast, the colors of the $3.8<z<5.0$ galaxies are 
shifted toward bluer values by H$\alpha$ emission (which lies in the 3.6$\mu$m bandpass).}
\end{figure}

We allow the differential extinction, E(B-V)$_{\rm{stars}}$, to range between 0.00 and 0.50 in steps of 0.02, and we limit 
the model ages to lie between 5 Myr and the age of the universe at the redshift of interest.    
The precise form of the dust attenuation curve is, of course, not known at $z>4$, but we consider 
the reddening law appropriate for local starbursts (e.g.,Meurer et al. 1999, Calzetti et al. 2000) 
and a steeper attenuation curve that is appropriate for the SMC (e.g.,Gordon \& Clayton 1998).   The 
latter appears to be appropriate for young galaxies ($<$100 Myr) at high redshift (Siana et al. 2008, Reddy et al. 
2010, 2012), a population which might become increasingly dominant at $z>4$.

\begin{figure*}
\epsscale{1.0}
\plotone{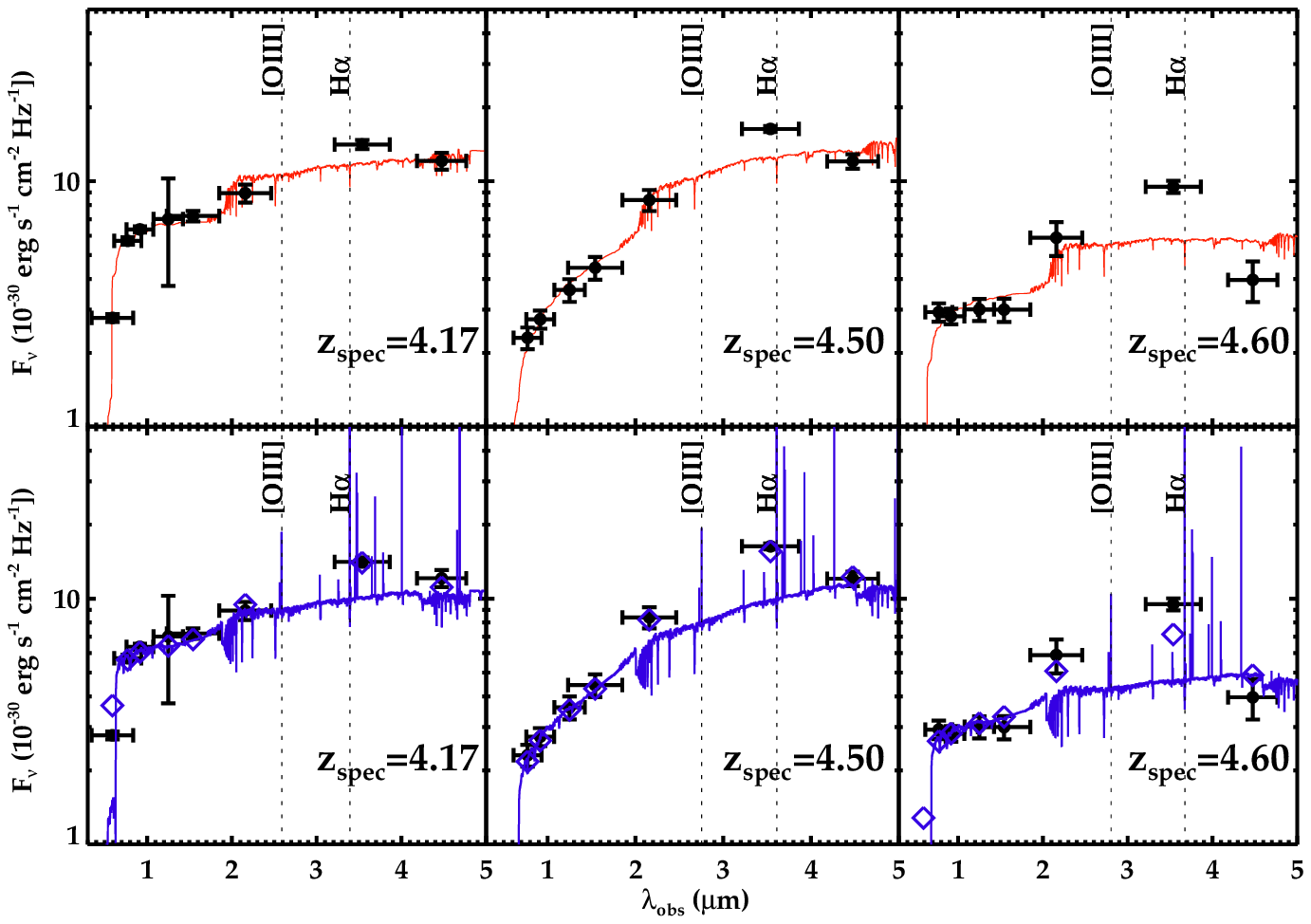}
\caption{SEDs of spectroscopically confirmed galaxies at $3.8<z<5.0$ fit using population synthesis models 
containing both stellar continuum (top row) and stellar+nebular emission (bottom row).
Many galaxies in this redshift range show blue [3.6]-[4.5] colours, with the 3.6$\mu$m flux significantly in excess of 
the best-fitting stellar continuum.   This flux excess is strongly suggestive of H$\alpha$ nebular line 
contamination.  Not surprisingly, models containing nebular emission provide significantly 
better fits to the observed photometry, as clearly indicated by the agreement between 
the synthetic (open blue diamonds) and observed (solid black circles) in the bottom row.   In the following, 
we will only consider objects for which the Balmer Break can be anchored by a significant ($S/N>5$) [4.5]
detection, removing fainter objects (like that in the right panel) for which an accurate flux excess is difficult  
to extract.
}
\end{figure*}

The relative extinction provided to stars and nebular emission is not definitively understood 
at high redshift.   Expectations from nearby galaxies suggest that the nebular gas is preferentially 
more extincted than the stellar continuum, as expected if the HII regions lie in dustier regions 
than the stars contributing to the integrated stellar continuum.   Based on observations of local star-forming 
galaxies and starbursts, Calzetti et al. (2000)  suggest that $A_{\rm{V,neb}}$ =  A$_{\rm{V, SED}}/0.44$.   
Whether or not this relationship holds at high redshift is unclear.    Some of the first studies of 
H$\alpha$ emission in star forming galaxies $z\simeq 2$ indicated that the nebular gas and stars 
might be equally attenuated (e.g., Erb et al. 2006, Reddy et al. 2010).    But more recently, 
new studies have emerged which support a factor $\simeq 2$ higher extinction toward HII regions 
(e.g., Forster-Schreiber et al. 2009, Onodera et al. 2010, Mancini et al. 2011, Wuyts et al. 2011), 
similar to that observed locally.  Clearly an improved understanding of how the relative distribution 
of stars and HII regions depends on age and mass would greatly benefit attempts to simultaneously 
fit stellar and nebular emission via population synthesis models.    In the nebular+stellar models presented 
in this paper, we will simply assume that A$_{\rm{V,SED}}$ = A$_{\rm{V,neb}}$.   

In the following, we will fit the data with both the nebular+stellar and stellar continuum models.     For the latter, 
the cleanest method is obtained by fitting the data excluding the [3.6] flux measurement, given that this band 
could be contaminated by H$\alpha$.  However, excluding this band means that only one filter is available to constrain 
the SED beyond the Balmer Break.   We therefore also fit the data using all 
photometric information, i.e. including the [3.6] measurement.   This is discussed further in \S3.2.  
For each galaxy, we compute the model age, normalization, and A$_{\rm{V}}$ which provide acceptable fits to the data.   
The normalization is then mapped to the star formation rate and stellar mass appropriate for the given template.  
We compute uncertainties on these parameters by bootstrap resampling the data within the allowed photometric uncertainties.   

\subsection{Nebular Line Strengths}

We infer H$\alpha$ emission line strengths in our sample of $3.8<z<5.0$ galaxies by 
comparing the observed flux density in the [3.6] bandpass to the flux density in that filter 
expected from stellar continuum alone.   We explore the method which produces the most 
accurate equivalent widths below. 

To verify the reliability of using the broadband flux excesses to derive emission line strengths, we 
examine the SEDs of a moderate redshift sample of nebular line emitters 
with spectroscopically-measured [OIII] line fluxes from WFC3 grism observations 
of the Hubble Ultra Deep Field (Trump et al. 2011).    We choose this sample rather than 
larger ground-based samples because use of 2D grism spectroscopy avoids uncertainties 
owing to slit losses.    To characterize the 
contribution of the emission lines to the broadband SEDs, we measure optical through mid-IR 
photometry using the UDF dataset (see McLure et al. 2011 for details) and perform 
population synthesis modelling following exactly the same procedure as described 
above.  [OIII] will either fall in the  J$_{125}$-band (at $1.3<z<1.8$) or 
H$_{160}$-band (at $1.8<z<2.3$).   We characterize the 
likely strength of the emission lines by comparing the observed broadband flux 
(in the contaminated filter) to the stellar continuum flux expected from the best-fitting 
population synthesis models.  We consider the 
stellar models with and without the contaminated bandpass included.  
In order to ensure a reliable measure of the stellar continuum in the vicinity of [OIII], 
we do not consider galaxies with strong emission lines in adjacent infrared filters 
(e.g., [OIII] in H$_{160}$ and H$\alpha$ in K$_s$) or those undetected in K$_s$-band.   

We focus our analysis on the four systems in this remaining subset for which the measured [OIII] flux is 
predicted to make a significant (e.g., $\gsim 4$\%) contribution to the broadband photometry.    In each of these systems, the 
contaminated filter reveals an excess with respect to the best-fitting stellar continuum model 
(Figure 2).   The fluxes required to produce the broadband excesses 
in the contaminated filter agree well with those measured spectroscopically (Figure 2), with the results 
from the two fitting methods typically bracketing the observed line flux.     While both methods produce 
remarkably good agreement, the line fluxes are slightly more accurate (average flux uncertainty of $\simeq 20$\%) 
when the stellar continuum is estimated from the fit to the SED including the contaminated filter.    When the 
contaminated filter is excluded from the fitting process, the inferred line flux is typically 1.5-2.0$\times$ greater 
than measured with the WFC3 grism.   Given  the SEDs are fairly poorly sampled in the  wavelength range 
where the continuum flux is required, it is conceivable that by excluding the contaminated filter, the fitting process 
will prefer redder models which fit the flux in the adjacent filters but underpredict the continuum in vicinity of the 
emission line of interest.   Regardless of the precise reason, it is clear from Fig. 2 that by considering the 
results from both methods, fairly accurate line strengths can be extracted from the photometry.

The results of this test therefore motivate use of the broadband flux excesses to infer line strengths in carefully-selected 
spectroscopic samples at higher redshifts where direct spectroscopic measurements of nebular line fluxes are not yet available.  
As a result of the higher redshifts (and the corresponding $1+z$~boost in observed equivalent width), we 
expect the nebular line contribution to broadband fluxes to be greater than typically observed at intermediate 
redshifts (e.g., Trump et al. 2011), allowing the flux excesses to more consistently stand out with respect to photometric 
uncertainties.    Based on the results in this section, we expect the the true line fluxes to be bracketed by our
two fitting methods, with the most accurate measurements obtained by fits to the entire SED.

\section{Results}

We now discuss the results derived from applying our technique to the SEDs of spectroscopically-confirmed 
galaxies in the redshift range $3.8<z<5.0$ over which H$\alpha$ may contaminate the IRAC 3.6$\mu$m filter.   
We compare the observed [3.6] flux densities to the stellar continuum expected from population synthesis 
models and use the results to infer an empirically-based H$\alpha$ equivalent width distribution (\S4.1).   
We discuss the possible redshift evolution of the nebular line strengths in \S4.2.   Using the empirically-derived 
equivalent width distributions, we examine how nebular emission affects the derived physical 
properties of $z>4$ galaxies (\S4.3).   

\subsection{Strength of Nebular Emission Lines}

We begin by comparing the [3.6]-[4.5] colour distribution for galaxies at $3.8<z<5.0$ with 
that for galaxies at $3.1<z<3.6$ (a redshift range over which the IRAC colours are uncontaminated 
by strong nebular emission). This should reveal the impact of H$\alpha$ emission on  
broadband fluxes at $z>3$.  We apply this test for 45 galaxies at $3.8<z<5.0$ with 
robust flux measurements in the 4.5$\mu$m filter which constrains the rest-optical stellar 
continuum.  The results  (Figure 3) point to a significant contribution 
from H$\alpha$.   The median [3.6]-[4.5] colour at $3.8<z<5.0$ is 0.33 mag bluer than the median 
value at $3.1<z<3.6$, consistent with expectations if H$\alpha$ pollutes the 3.6$\mu$m filter in the 
higher redshift bin.   Note the slightly red [3.6]-[4.5] colors of the $3.1<z<3.6$ sample are exactly what is 
expected for moderately reddened ($E[B-V]\simeq 0.1$) galaxies with a constant star formation history 
and luminosity-weighted ages of 100 Myr.   A Kolmogorov-Smirnoff test demonstrates with confidence that these two color 
distributions are distinct (KS statistic of D=0.54 with an associated probability by chance of 8$\times$10$^{-8}$).   
 We also consider whether the change in color might be due to photometric scatter 
from increased photometric error in the higher redshift bin.   We test this by randomly 
perturbing the $3.1<z<3.6$ [3.6]-[4.5] colour distribution according to the IRAC flux errors of the 
$3.8<z<5.0$ sample.    While this can slightly broaden the width of the colour distribution, it does 
not shift the median colour to bluer values as observed. 

While the most natural interpretation of the systematic offset is the presence of H$\alpha$ in the 
[3.6] filter, it is conceivable that other effects could contribute.   For example, one 
might expect that a systematic offset in [3.6]-[4.5] colours might arise from the slightly 
different rest-frame wavelengths sampled and the (potentially) younger ages in the 
higher redshift bin.   Examination of population synthesis models indicates that intrinsic galaxy 
evolution is not likely to dominate the shift in [3.6]-[4.5] colors.   Given the median reddening and 
ages inferred for the $3.1<z<3.6$ and $3.8<z<5.0$ spectroscopic samples, we would expect to see 
[3.6]-$[4.5] \simeq 0.1$.   This is similar to that observed at $3.1<z<3.6$, but significantly redder than 
that observed in the $3.8<z<5.0$ redshift range with H$\alpha$ contamination.  We therefore conclude that 
nebular contamination is likely the dominant cause of the differences in the [3.6]-[4.5] colours.

A particularly convincing verification of the above statistical test is the fact that we can directly see
evidence of strong nebular emission in individual SEDs (Figure 4). Clearly in these examples the flux 
in the 3.6$\mu$m filter is not only in excess of that at 4.5$\mu$m but is also significantly in excess of 
the stellar continuum of the best-fitting population synthesis models.   The SEDs of galaxies in this 
redshift range are (not surprisingly) typically better fit by models including nebular emission (blue lines in 
bottom panel cf. red lines in top panel for stellar continuum models in Figure 4).   

To estimate the strength of H$\alpha$, we compute the amount by which the observed 3.6$\mu$m 
flux exceeds the predicted stellar continuum flux.   We define the 3.6$\mu$m excess, $\Delta$[3.6] as 
the difference between the [3.6] magnitude expected from stellar 
continuum models which fit the SED and the observed [3.6] magnitude.    Positive values indicate that the observed flux is greater than can 
be accommodated by stellar continuum.   This test requires an accurate measure of the stellar continuum in the rest-optical.  
Again we limit our sample to those galaxies with confident [4.5] detections, as this filter 
(devoid of strong emission lines) is necessary to anchor the population synthesis models beyond the 
Balmer Break.   

The distribution of 3.6$\mu$m magnitude excesses in our spectroscopic sample 
(Figure 5) reveals that 96\% of galaxies are observed to be brighter at [3.6] then predicted from the 
best-fitting stellar continuum models.   The median excess, 0.27 mag, suggests that the typical rest-frame 
emission line equivalent width in the 3.6$\mu$m filter at $3.8<z<5.0$ is 360-450~\AA.   Emission 
lines therefore contribute nearly 30\% of the observed [3.6] broadband photometry. 
These values are derived from stellar continuum model fits that include the contaminated 3.6$\mu$m filter 
in the modelling.    As we demonstrated in \S3, this method produces the 
most accurate flux estimates.  If the contaminated filter is excluded from the modelling procedure, the 
inferred continuum level is typically reduced, resulting in a slightly larger median [3.6] excess (0.37 mag) and 
total equivalent widths (520-650~\AA).   Note these represent the equivalent width of all emission lines 
in the [3.6] filter.   We derive the fractional contribution of H$\alpha$ below.  
Based on the discussion in \S3.2, it is likely that these measurements bracket the range of mean 
nebular emission line strengths in  $3.8<z<5.0$ galaxies.   Reassuringly, this EW range is consistent with that 
required to explain the 0.33 mag offset in median [3.6]-[4.5] colors of  $3.1<z<3.6$ and $3.8<z<5.0$ galaxies in 
Figure 3.

To estimate the {\it equivalent width distribution} of emission lines contaminating the IRAC 3.6$\mu$m filter, 
we need to admit a range of equivalent widths to reproduce the observed 3.6$\mu$m 
photometric excess distribution of Figure 5a.   We assume that equivalent widths are distributed 
in a log-normal fashion, similar to that seen from H$\alpha$ emission locally and at moderate redshifts 
(e.g., Lee et al. 2007, Ly et al. 2011).   We consider a large grid spanning a range of $\sigma$ and $\mu$, 
the width and mean of the equivalent width distribution.  We translate each equivalent width into a 3.6$\mu$m 
excess, applying a photometric scatter of 20\% (a conservative estimate for the average 3.6$\mu$m magnitude error) 
and compute the flux excess distribution expected from the input equivalent width distribution.   We find that     
the observed flux excess distribution is well-fit by an equivalent width distribution with $\sigma=0.25$ and 
$<\rm{log_{10} (W_{[3.6]}/\AA)}>$=2.57 (Figure 5b).   This equivalent width will surely be dominated by 
H$\alpha$ emission, but other emission lines ([SII], [NII]) may of course contribute.  The contribution of other lines 
will depend on the physical properties (e.g., metallicity) of the galaxies.   Our sub-solar (0.2 Z$_\odot$)
metallicity models indicate that H$\alpha$ should contribute $\simeq 76$\% of the observed equivalent width.  
In this case, the typical H$\alpha$ EW at $3.8<z<5.0$ is  $<\rm{log_{10} (W_{H\alpha}/\AA)}>$=2.45.   With these 
assumptions, if the [3.6] filter is excluded from the fitting, we find $<\rm{log_{10} (W_{H\alpha}/\AA)}>$=2.61.   As above, 
we adopt this as an upper bound to the average H$\alpha$ EW.   

The level of H$\alpha$ emission quoted above is actually very reasonable given the typical properties 
of $z\simeq 4-5$ LBGs (e.g., Stark et al. 2009, Gonzalez et al. 2011a).  For constant star formation, ionizing 
photon escape fraction of 0.2, and ages of $\simeq 100-250$ Myr, we would expect H$\alpha$ EWs to be 
$\simeq 200-300$~\AA, similar to the range we infer.     So the observation of [3.6] excesses of 0.2-0.3 mag 
relative to stellar continuum (Figure 3)  is exactly what we would expect given the shape of the overall SEDs.   
Indeed, the absence of any nebular contamination at $3.8<z<5.0$ would have been a far more 
surprising finding.  

\begin{figure*}
\epsscale{0.5}
\plotone{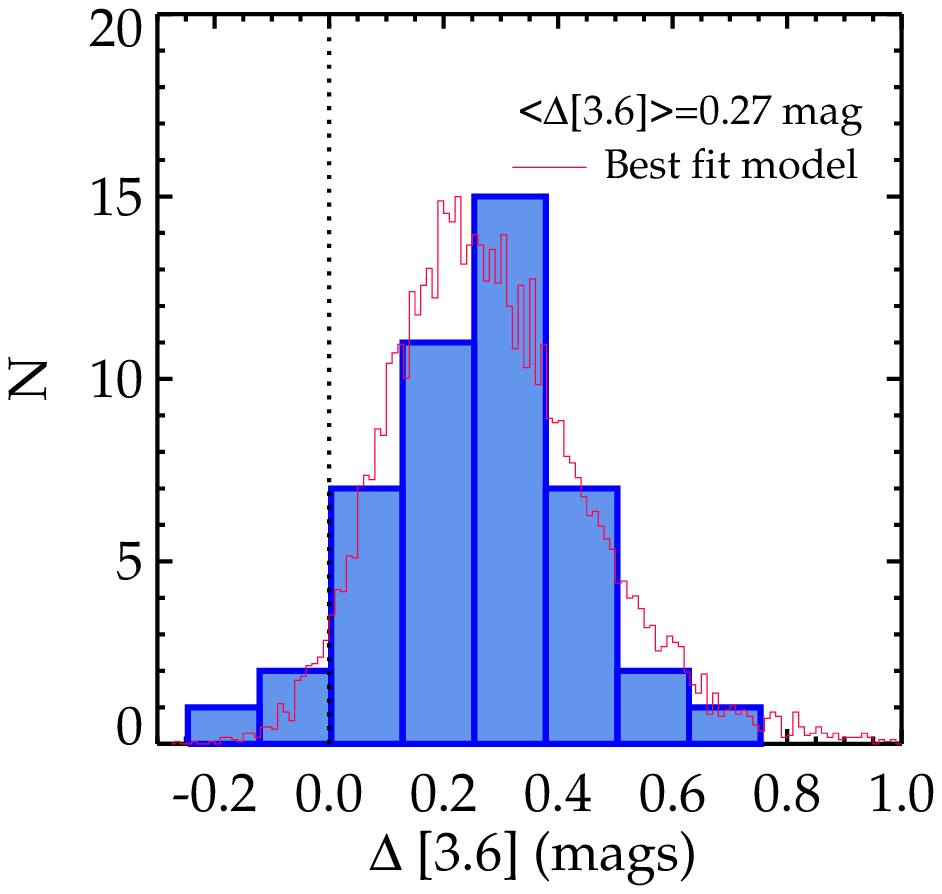}\plotone{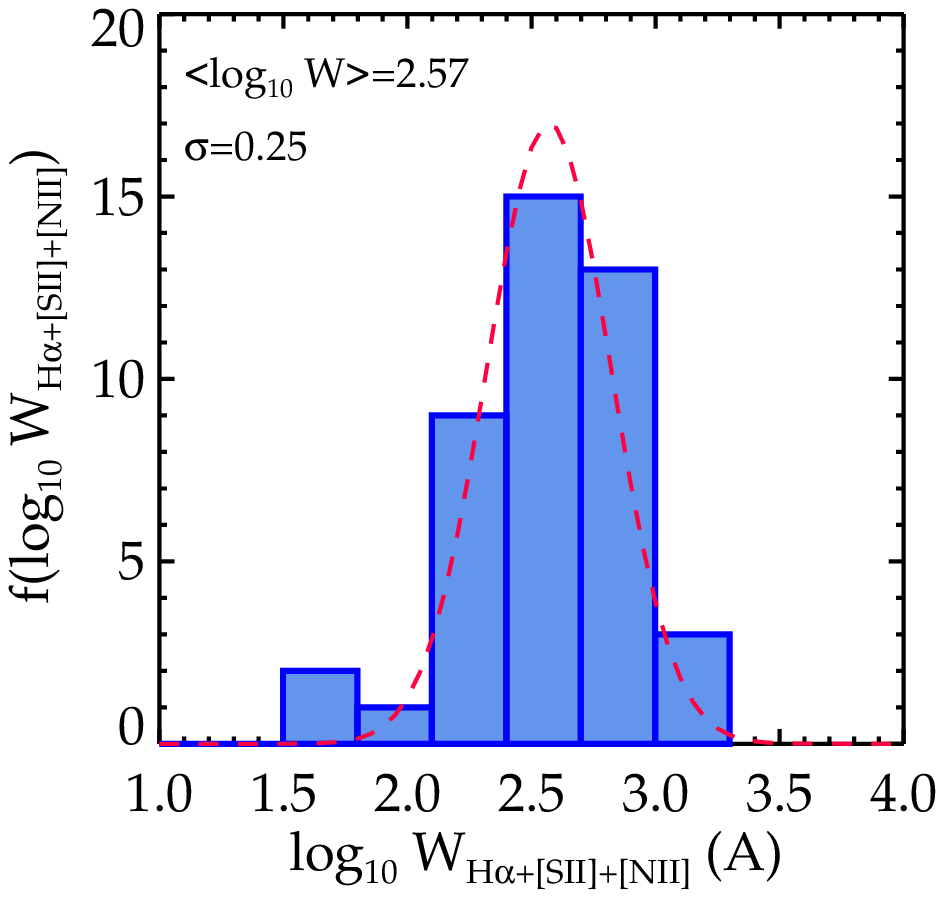}
\caption{
{\it Left:} The distribution of 3.6$\mu$m magnitude excesses ($\Delta$[3.6]) in our $3.8<z<5.0$ galaxy sample.   The 
magnitude excess is defined as the difference between the [3.6] magnitude inferred from the stellar continuum of 
the best-fitting population synthesis model and the [3.6] magnitude observed with {\it Spitzer}/IRAC.   The positive $\Delta$[3.6] 
values exhibited by our sample indicate that the stellar continuum is unable to account for the observed flux 
in the 3.6 $\mu$m filter.    The magnitude excess distribution is well fit by a log-normal equivalent width 
distribution with $<\rm{log_{10} W}>$=2.57 and $\sigma=0.25$ (red curve). {\it Right:}  Distribution of equivalent 
widths required to reproduce the observed flux excesses (blue histogram) compared to the functional form we 
adopt for the equivalent width distribution (red curve).  }
\end{figure*}

\subsection{Evolution of Nebular EW Distribution}

Before evaluating how nebular emission affects the derived stellar masses at $4<z<7$, 
it is interesting to consider whether the nebular equivalent width distribution is likely to evolve with
redshift.   Fumagalli et al. (2012) recently examined the evolution of  H$\alpha$ equivalent 
widths at lower redshifts, finding that the evolution could be fit by a power law 
$\propto(1+z)^{1.8}$  for galaxies with 10$^{\rm{10}}$ M$_\odot$ in stellar 
mass (Figure 6).    While our sample size is too modest to permit a detailed comparison 
with this trend at $z\gsim 4$, it is of interest to consider how the EWs we derived in \S4.1 compare 
to those at lower redshift.  We note that the H$\alpha$ equivalent widths presented in Fumagalli et al. (2012) include the 
contribution of [NII], while the nebular lines strengths we infer from photometric excesses 
include the contribution of all emission lines contaminating the [3.6] filter.   Using our 0.2 Z$_\odot$  
population synthesis models, we estimate that 82\% of the  equivalent width inferred from 
[3.6] photometric excesses arises from H$\alpha$ and [NII].    Note that this is slightly larger than the 
percentage estimated in the previous section owing to the addition of [NII] to the calculation.

Applying this factor to the mean equivalent widths presented in \S4.1, we compare the H$\alpha$+[NII] 
equivalent widths at $3.8<z<5.0$ to those at $z\lsim 2$ (Figure 6).   It is clear that the line strengths derived 
at $3.8<z<5.0$ are consistent with a continued increase 
in the H$\alpha$ EW in the range $2\lsim z\lsim 5$.    While determination of the exact rate of increase 
is beyond the scope of this work, we note that the H$\alpha$ EWs at $3.8<z<5.0$  lie in the range 
expected by simple extrapolations of the power laws derived in Fumagalli et al. (2012).  This 
increase in H$\alpha$ EW over $2<z<5$, albeit tentative in nature, is supportive of an increase 
in  the sSFR over $z\gsim 2$, consistent with more recent derivations (e.g., Bouwens et al. 2012).

Given these results, it is reasonable to expect nebular lines to be even stronger at $z\simeq 6-7$.   
Unfortunately, with both IRAC filters contaminated by strong emission lines at these redshifts (Figure 1), 
we do not have a direct method of estimating the nebular line contamination in this regime.   We 
thus will consider two cases in the following sections.   First, we assume conservatively that the nebular line 
strengths remain fixed at the values derived at $3.8<z<5.0$.   While a fixed EW might seem unlikely in light 
of the power law evolution at lower redshifts, we note that the rate of increase in the EW might 
slow if star formation histories transition into a phase of rapidly rising star formation rates (e.g., Finlator et al. 2011) at $5<z<7$.   As a modest upper limit, we consider the case whereby the nebular equivalent widths continue to increase following a (1+z)$^{1.8}$ power law.   We conservatively adopt the mean EW of the 
fitting method including the contaminated [3.6] filter as the $3.8<z<5.0$
reference value for this upper bound.   

\begin{figure}
\epsscale{1.0}
\plotone{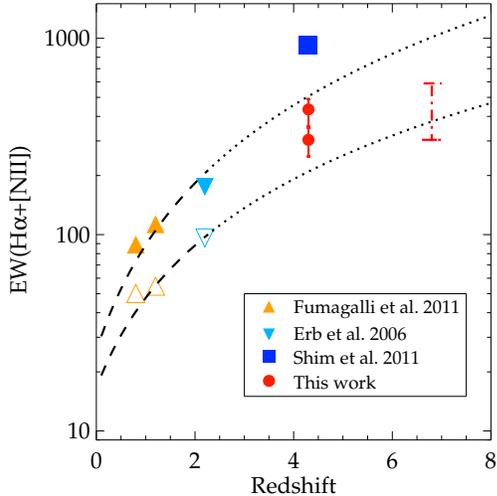}
\caption{Evolution of the mean H$\alpha$+[NII] EW with redshift.   The values at $z<4$ are as compiled in 
Fumagalli et al. (2012) for star forming galaxies with stellar mass in the range log$_{\rm{10}}$M$_\star$=10.0-10.5. 
The dashed lines show the power law that  Fumagalli et al (2012) fit to the EW
evolution at $z\lsim 2$ while the dotted lines show an extrapolation of this power law to $z\gsim 2$.  The 
 open symbols denote the EWs appropriate for entire galaxy population in Fumagalli et al. (2012), 
 while the solid symbols denote estimates for the star-forming subset.
The red circles show EWs inferred from the photometric excess method in 
this paper, with the lower value arising from SED fits including all filters, and the upper value derived from 
fits excluding the contaminated [3.6] filter (see \S3.2 for details).   The range of EWs illustrated at $z\simeq 6-7$    
represents nebular line strengths that we apply to SEDs in \S4.3, with the lower limit 
assuming that nebular line strengths remain fixed with increasing redshift and the upper limit assuming they
follow a (1+z)$^{1.8}$ power law.
}
\end{figure}

\begin{figure*}
\epsscale{1.1}
\plotone{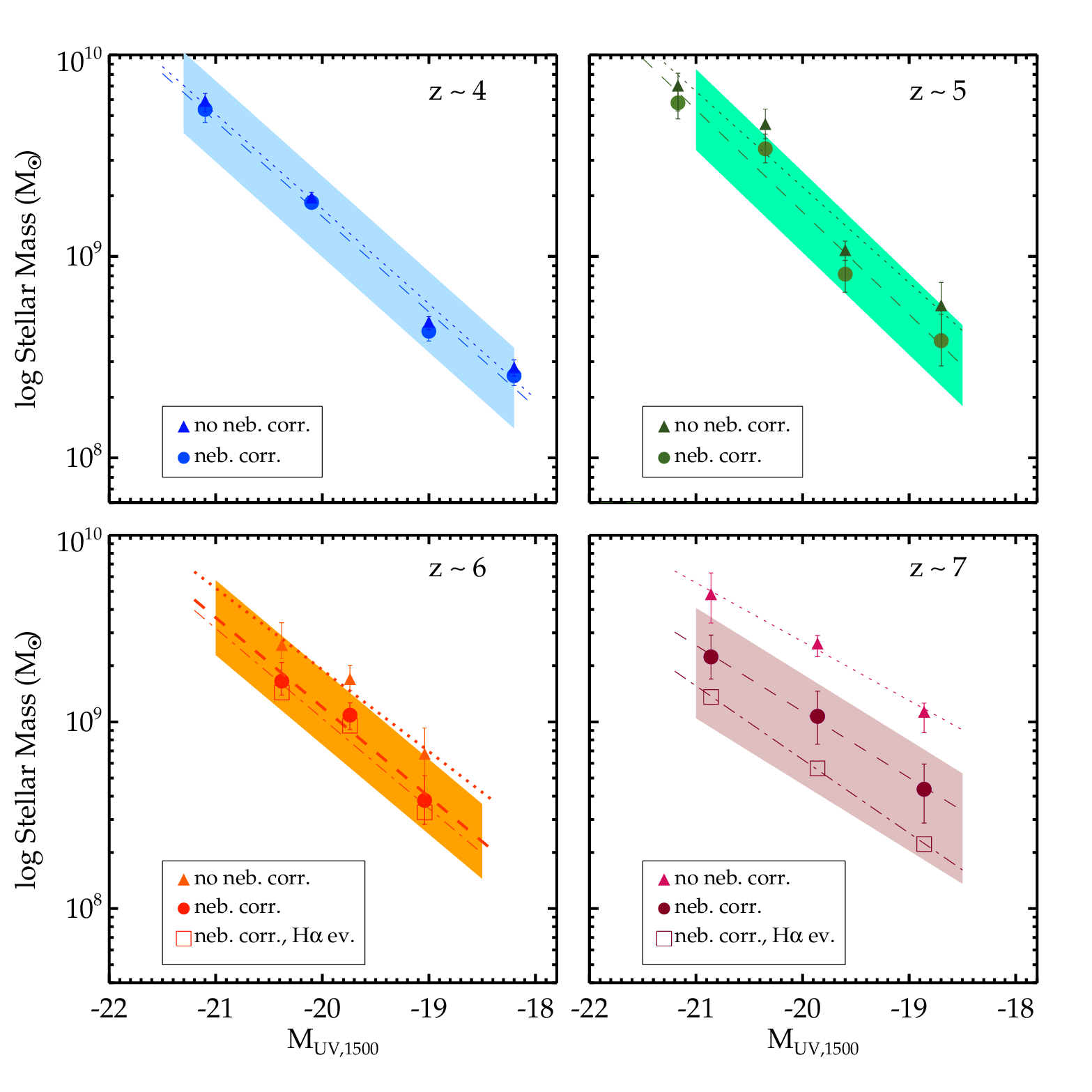}
\caption{The impact of nebular emission on stellar mass derived using empirical determination of H$\alpha$ 
EW distribution at $3.8<z<5.0$.  Solid circles show the log M$_\star$ - M$_{\rm{UV,1500}}$ relationship 
corrected for nebular contamination of broadband fluxes following the procedure discussed in \S4.2.  
The dashed line gives the best linear fit to these data points.   Solid triangles show the relationship with no 
correction for nebular emission assuming that the stellar continuum dominates the broadband flux.  
Without accounting for nebular emission, the log M$_\star$ - M$_{\rm{UV,1500}}$ relationship does not 
evolve much with redshift.  However, incorporating nebular corrections, the normalization {\it decreases} by 
1.4-2.5$\times$ over $4<z<7$.   The open squares correspond to nebular corrections derived assuming that the EW of nebular emission 
increases with redshift following the power law shown in Figure 6.}
\end{figure*}

\subsection{Effect on $z\gsim 4$ Stellar Masses}

In \S4.1, we demonstrated that strong nebular line emission lines make a 
significant contribution to the broadband flux measurements at $z>3$.  If these lines are not 
accounted for in population synthesis modelling, the rest-optical stellar continuum (and 
thus the inferred stellar mass and age) will clearly be overestimated (e.g., Schaerer \& de 
Barros 2010).   In principle, these issues can be addressed through nebular+stellar population 
synthesis models described in \S3.   The drawback of this approach is that 
the 'appropriate' flux from nebular emission for any given model is very uncertain,  depending 
not only on the escape fraction of ionizing radiation but also on the reddening law for the 
nebular gas, both of which are not known at $z>3$.   

Here we have attempted to account for these shortcomings using a method that relies on our 
empirically-derived nebular line equivalent width distribution.   For each SED we wish 
to fit, we draw a large number ($N\simeq 10^4$) of H$\alpha$ emission line equivalent widths from the 
distribution we derived in \S4.1 (e.g., Figure 5b).    The contribution from [OIII] and H$\beta$ (which 
contaminate the {\it Spitzer}/IRAC filters at $z\gsim 5$) is obtained by scaling the H$\alpha$ EW by 
1.7-2.0$\times$ (the exact value chosen at random from a uniform distribution), consistent with the flux ratios 
observed in sub-solar galaxies at $z\simeq 2-3$ (e.g., Hainline et al. 2009, Bian et al. 2010, Erb et al. 2010, 
Richard et al. 2011) and those predicted by our nebular+stellar population synthesis models.
If the galaxy's redshift places any of these strong lines in the broadband filters we are fitting, we subtract the 
predicted nebular flux from the photometry.  

In order to evaluate how the average properties of the various $z\simeq 4-7$ dropout 
populations are affected by nebular emission, we consider {\it composite 
SEDs} for the B, V, $i'$, and $z$-band dropout populations (e.g., Stark et al. 2009, Labb{\'e} et al. 2010b, 
Gonzalez et al. 2011b) binned by rest-UV magnitude.  For the purposes of this section, we limit our 
analysis to the most recent determinations of the composite SEDs (Labb{\'e} et al. 2010b, 
Gonzalez et al. 2011b), both of which take advantage of WFC3 photometry in the UDF and GOODS 
fields.   Since the effect of nebular emission on derived physical 
properties is clearly redshift-dependent (c.f. Figure 1),  we must account for the distribution of redshifts within 
a particular dropout sample. Thus for each realization of the nebular line EW distribution, we also 
select a redshift from the expected photometric redshift distribution of the dropout population 
under consideration (see e.g., Bouwens et al. 2012).  We fit these realizations of the composite SEDs 
using the stellar continuum models described in \S3.  Physical properties are determined in a similar 
manner to that described in \S3, with uncertainties derived from the 1$\sigma$ intervals of the large 
number of realizations of the equivalent width distribution.   

\begin{figure*}
\epsscale{0.52}
\plotone{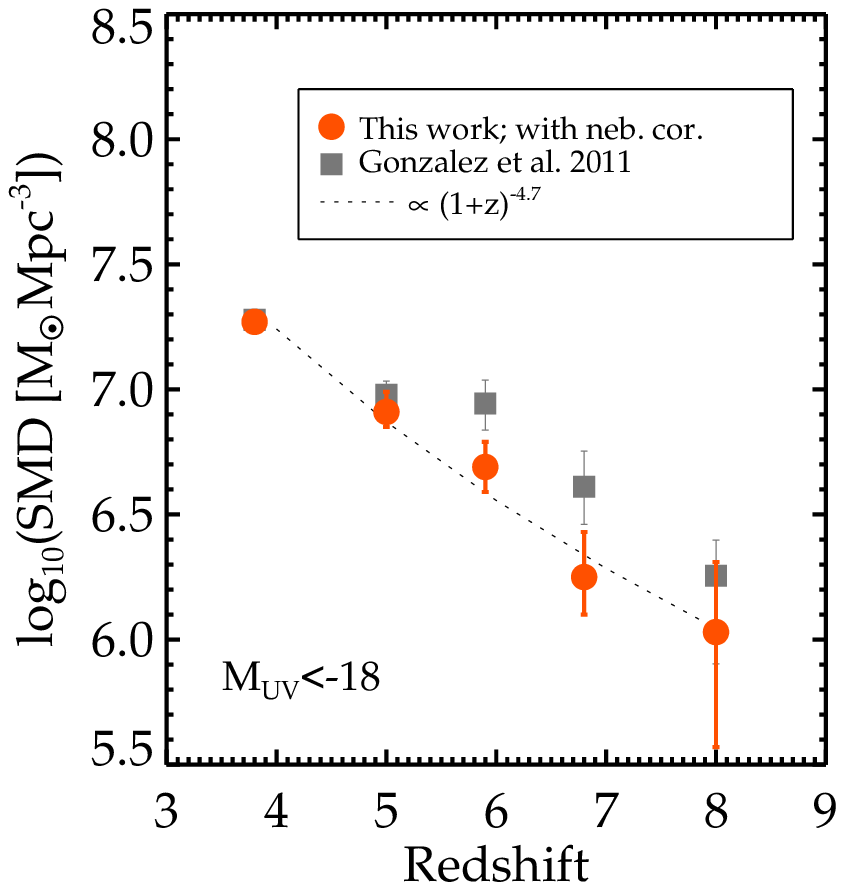}\plotone{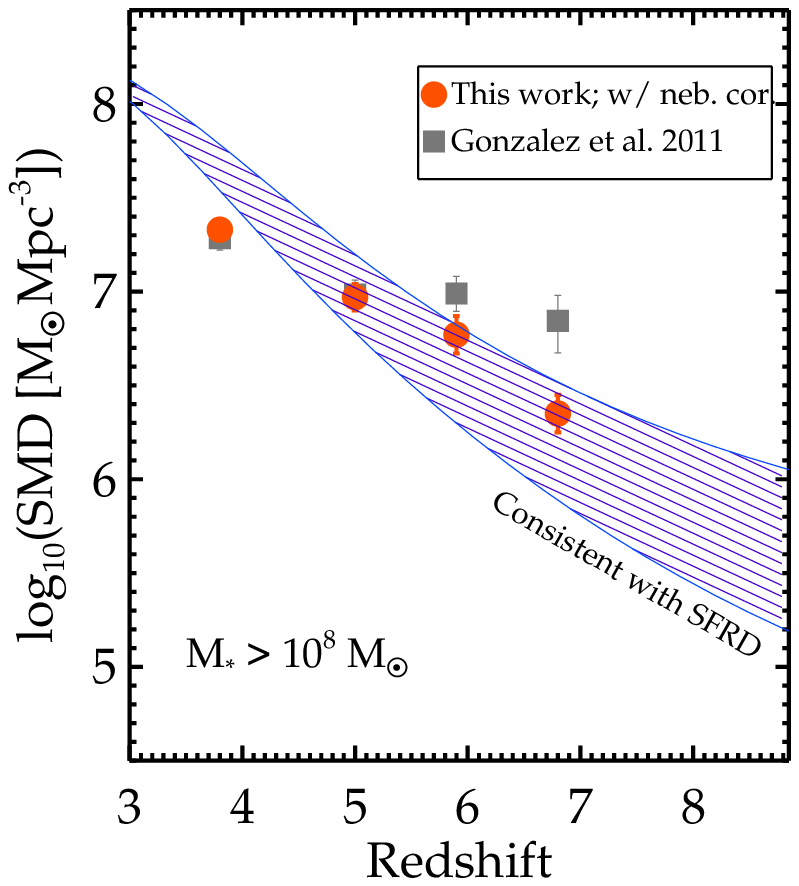}
\caption{Evolution in the Stellar Mass Density: The mass density is computed by integrating the stellar mass 
function to a fixed UV luminosity limit (left panel) and fixed stellar mass limit (right panel).  The main advance
with respect to earlier work is  the inclusion of corrections for nebular emission contamination of {\it Spitzer}/IRAC filters. 
These are computed using the nebular EW distribution derived from our spectroscopic sample in \S4.1 and assume the 
H$\alpha$ EW distribution continues to evolve as a power law at $z\gsim 5$. 
The stellar mass functions are  derived by combining the nebular-corrected log M$_\star$-L$_{\rm{UV}}$ relation 
with the UV luminosity functions presented in Bouwens et al. 2012.  We assume that the scatter in the 
log M$_\star$-L$_{\rm{UV}}$ relation is  $\sigma\simeq 0.5$, consistent with previous studies 
(e.g., Gonzalez et al. 2011a).    The blue swath in the right panel shows the stellar mass density implied 
by the evolving star formation rate density, computed by integrating the UV luminosity functions (see Robertson et al. 2010 
for details).}
\end{figure*}

The impact of nebular emission is clearly seen in the log M$_\star$ - M$_{\rm{UV}}$  scaling relations 
shown in Figure 7.  These are determined for each dropout population 
with and without the nebular correction from the composite SEDs discussed above.  The absolute 
magnitudes are unchanged in this analysis, so the changes shown are due only to the 
contamination of broadband light by nebular emission lines.  As expected, the impact of nebular emission 
is strongest in the range $5<z<7$ where nebular lines contaminate both IRAC filters.  In contrast, the nebular 
correction is less severe for the B-drop ($z\simeq 4$) population, since the Spitzer/IRAC 4.5$\mu$m filter 
is devoid of strong emission lines throughout the redshift range covered by B-drops (Figure 1).    

Assuming the nebular line EW distribution at $z>5$ remains identical to that determined at $3.8<z<5.0$, we 
find that the average stellar masses  are reduced by factors of $\times$1.1, 1.3, 1.6, and 2.4
for the dropout populations centered at $z\simeq \rm{4,5,6,~and~7}$ respectively.   If the nebular line 
strengths increase with redshift at $z\gsim 5$ following a (1+z)$^{1.8}$ power law (Figure 6; Fumagalli et al. 2012), 
the typical stellar masses  are reduced by $\times$1.9 and 4.4 at $z\simeq 6$ and 7 respectively.  
We emphasize that these represent average corrections applicable to the B, V, $i'$, and $z$-band dropout populations.   
Individual galaxies throughout this redshift range will of course be affected differently.

To summarize, we have used the equivalent width distributions derived in \S4.1 to compute the 
likely contribution of nebular emission to broadband photometry.   We find that stellar masses at 
$z\gsim 6$ need to be revised downward by 2-4$\times$, with the precise correction depending on whether 
the EW of H$\alpha$ and [OIII] emission continue to increase with redshift beyond $z\simeq5$.   
This result has an important effect on the log M$_\star$ - M$_{\rm{UV}}$ scaling relation, which previously was thought to be 
largely constant with redshift at $z>4$ (e.g., Stark et al. 2009, Gonzalez et al. 2011, McLure et al. 2011).   
It is actually clearly evident in the uncorrected log M$_\star$ - M$_{\rm{UV}}$ relations presented 
in Figure 7 that without nebular corrections, the M$_\star$/L$_{\rm{UV}}$ ratios implied by the 
composite SEDs {\it increase} with redshift.  Our analysis indicates that this finding is likely to be an
artifact of nebular contamination.   After correcting for line emission, we  demonstrate that the M$_\star$/L$_{\rm{UV}}$ ratios 
are likely to decrease by $\times$1.4-2.5 with redshift over $4<z<7$.  This result has important 
implications for derivation of the stellar mass density (\S5.1) and specific star formation 
rate evolution (\S5.2) discussed below.

\subsection{Effect of Nebular Continuum Emission}

Nebular continuum emission can also contribute to the observed broadband flux density.   
Most importantly, the addition of nebular continuum reddens the intrinsic spectrum at young ages, 
thereby requiring less dust extinction to reproduce the observed colors.  This can lead to a 
reduction in the derived star formation rates, in addition to the reduction in the stellar masses 
discussed earlier.  The effects of nebular continuum can be seen in the nebular+stellar models, 
which typically display spectral edges in the vicinity of the Balmer/4000~\AA~ break 
(bottom panel of Figure 4).   Whether significant nebular continuum is actually contributing to the spectra 
of high-z galaxies is unclear.   Without better sampled SEDs (from e.g., 
medium-band near-IR photometry) that might probe such spectral edges, it is difficult to  verify the presence of  
nebular continuum, as we are able to with nebular emission lines.  Nevertheless if emission lines are 
very strong, it is likely that there is also a significant contribution from nebular continuum emission.  

To quantify the likely impact of nebular continuum on the derived physical properties, we 
examine how the inferred dust reddening (and SFR) are affected when the nebular continuum 
is added.   To do so, we compare the dust attenuation necessary to reproduce a fixed UV continuum 
slope for models with and without nebular continuum emission included.  Assuming a Calzetti 
reddening law, we calculate the the dust attenuation that reproduces a UV slope with  
$\beta=-1.5$ as a function of model age.   Owing to the redder intrinsic slopes, the inferred 
dust attenuation is reduced for the nebular+stellar models.   The effect is most pronounced at 
the youngest ages, with the inferred attenuation up to 20\% lower for nebular+stellar models 
with ages $<$30 Myr.   As a result of the reduced attenuation, a 
lower normalization is required to match the observed flux density, bringing down the inferred 
star formation rates by up to the $\simeq 20$\% level for the youngest systems.  In practice, 
the impact of nebular continuum is more complicated and non-trivial to predict, depending 
strongly on the shape (i.e. age) of the observed SED.   Given these uncertainties, our analysis 
in the following sections will focus on how physical properties are affected by nebular emission lines.

\section{Discussion}

In the previous section, we used the broadband SEDs of a large sample of spectroscopically-confirmed 
galaxies to infer the distribution of nebular line strengths in UV-selected galaxies at $3.8<z<5.0$.    We 
showed that the stellar masses inferred from population synthesis modelling are reduced  
at $z>5$ when the contribution of these lines to broadband flux densities is removed.    In this section, we 
consider the implications of these results for our current picture of early mass assembly.

\subsection{Stellar Mass Density at $z>3$}

In \S4.2, we quantified the extent to which the stellar masses of $z>3$ galaxies are 
affected by nebular emission.  Here, we seek to utilize these results to estimate   
the stellar mass density (SMD) evolution at $z>3$.      To derive the mass densities, we combine the 
log M$_\star$ - M$_{\rm{UV}}$ relationship with UV luminosity functions (LFs) in a manner mostly 
similar to that outlined in Gonzalez et al. (2011).     Briefly, we extract a large number 
($N\simeq 10^5$) of luminosities from the measured UV LFs (e.g., Bouwens et al. 2011). 
We convert these luminosities to stellar masses using the 
log M$_\star$ - M$_{\rm{UV}}$ relationship, and an estimate of the scatter about the median.  Whereas 
earlier studies held the log M$_\star$ - M$_{\rm{UV}}$ relationship fixed with redshift at $z\gsim 4$, the strong 
redshift dependence of nebular contamination (Figure 1) forces us to reconsider the evolution of 
M$_\star$/L$_{\rm{UV}}$ ratios with redshift.

We compute the slope and normalization of the $z\simeq 4$ log M$_\star$ - M$_{\rm{UV}}$ relationship using 
the large sample of LBGs discussed in Stark et al. (2009).    For simplicity, we assume  the slope 
remains constant at $z\gsim 4$ and consider only evolution in the normalization of the relationship.  To 
compute the zero-points of the log M$_\star$ - M$_{\rm{UV}}$ 
relation at $z\simeq 5$, 6, and 7, we adjust the measured $z\simeq 4$ relation to account for the relative normalisation of the 
nebular-corrected  log M$_\star$ - M$_{\rm{UV}}$ relationships shown in Figure 7.    To obtain a tentative  
estimate of the $z\simeq 8$ stellar mass density, we apply the $z\simeq 7$ log M$_\star$ - M$_{\rm{UV}}$ 
relationship to the $z\simeq 8$ UV LF.    In all cases, we use 
the nebular corrections derived assuming an evolving H$\alpha$ EW distribution (see Figure 7), but 
we also discuss how these results would change if the EW distribution remains fixed at $z>5$.

In addition to measurement of the log M$_\star$ - M$_{\rm{UV}}$ relationship, accurate determinations 
of the dispersion are necessary to account for low luminosity galaxies with large M$_\star$ /L$_{\rm{UV}}$ ratios.   
If scatter is not accounted for, the mass functions will be incomplete and mass densities (above a fixed mass limit) 
will be underestimated. While a measurement of the observed scatter at $z\simeq 4$ (0.5 dex) was made in 
Gonzalez et al. (2011a), the intrinsic scatter is likely lower due to systematic uncertainties in the modelling as well as the effects of 
nebular contamination.   In the following, we assume the intrinsic scatter is in the range 0.2-0.5 dex.

We focus first on the UV luminosity limited measure of the SMD, considering only those galaxies with 
luminosities greater than M$_{\rm{UV}}=-18$.  The results are presented in Figure 8.    As discussed in 
earlier sections, if nebular emission is not accounted for in the modelling, the inferred M$_\star$/L$_{\rm{UV}}$ 
ratios actually increase moderately with redshift, leading to artificially high stellar mass densities at $z\gsim 5$.  
The inclusion of nebular emission reduces the mass density at $z\simeq 7$ by up to 4$\times$, while having little effect at $z\simeq 4$.    
As a result of these redshift-dependent corrections, the evolution in the SMD is fit by a steeper power 
law ([1+z]$^{-4.7}$)  than reported previously.   

We also consider the SMD in galaxies with stellar masses in excess of 10$^{8}$ M$_\odot$.   For 
consistency with earlier measurements of the SMD, the measurements we present in Figure 8 assume 
that the  log M$_\star$ - M$_{\rm{UV}}$ relationship has an intrinsic dispersion of 0.5 dex.   If the scatter 
is instead only 0.2 dex, for example, we would find stellar mass densities that are $\simeq 1.6-2.0\times$ 
lower.   

While previous estimates of the $6\lsim z\lsim 7$  SMD appeared broadly consistent with the integral of 
the $z\gsim 7$ SFRD (after an appropriate correction 
for stellar mass loss and recycling), the SMD appeared to be on the high end of the range 
implied by the SFRD (e.g., Robertson et al. 2010).   We compare our revised SMD to those 
implied by the SFRD in the right panel of Figure 8.     The mass density implied by the SFRD is 
calculated in a similar manner as Robertson et al. (2010), updated to include the latest measurements 
of the UVLF (Bouwens et al. 2011).    With appropriate corrections for nebular emission, the 
mass densities now appear in better agreement with the integrated SFRD.

The stellar mass density provides a useful integral constraint on earlier star formation.    {\it Spitzer} 
observations of galaxies at $z\simeq 6-7$ therefore offer a valuable measure   
of the likely ionizing output of galaxies at $7\lsim z\lsim 15$.   Therefore as measurements of the stellar mass 
density become more reliable, they will offer insight into the contribution of galaxies to reionization, 
complementing inferences from the UV luminosity function.   

We have demonstrated in this paper that 
corrections for nebular emission are a crucial aspect of obtaining a reliable census of stellar mass 
in the early universe.    In light of the reduced mass density required by nebular contamination, we 
re-consider the ability of galaxies to achieve reionization by $z\gsim 6$, updating the calculation presented 
in Robertson et al. (2010).    Given the consistency with the SFRD (Figure 8), these results are not 
surprisingly similar to inferences obtained from the UV LF (e.g., \citealt{Bouwens12a,Kuhlen12}).    
The UV output implied by the mass density is in principle sufficient to achieve reionization by 
$z\simeq 6-8$ but struggles to account for the optical depth to electron scattering implied by WMAP 
(e.g., Larson et al. 2011).
 
Understanding this photon shortfall will require improved knowledge of how much star formation 
occurs beyond $z\simeq 10$.   While direct detection of $z\gsim 10$ galaxies will likely have to 
wait until JWST,  {\it Spitzer} offers a unique means of progress in the coming years.   By obtaining 
 stellar mass estimates for the emerging samples of $z\simeq 9-10$ galaxies (e.g Bouwens 
et al. 2011b, Zheng et al. 2012), it will be possible to obtain some of the first constraints on the 
contribution of galaxies to the cosmic ionisation history beyond $z\simeq 10$.   

\subsection{sSFR Evolution}

\begin{figure*}
\epsscale{0.5}
\plotone{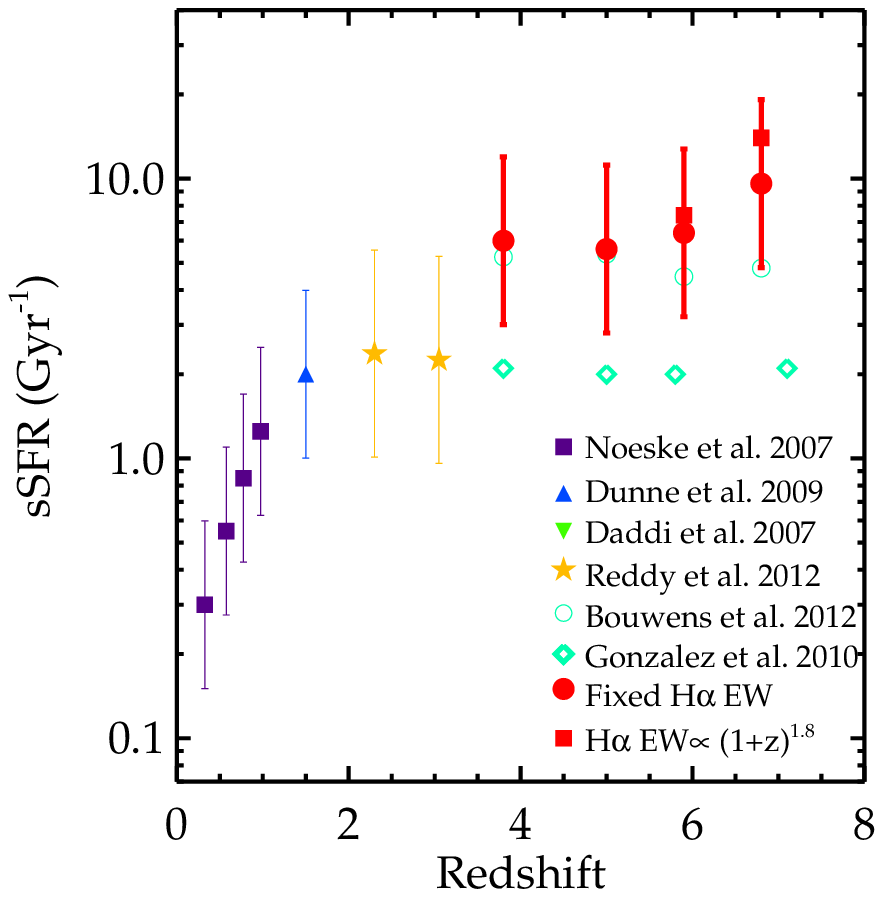}\plotone{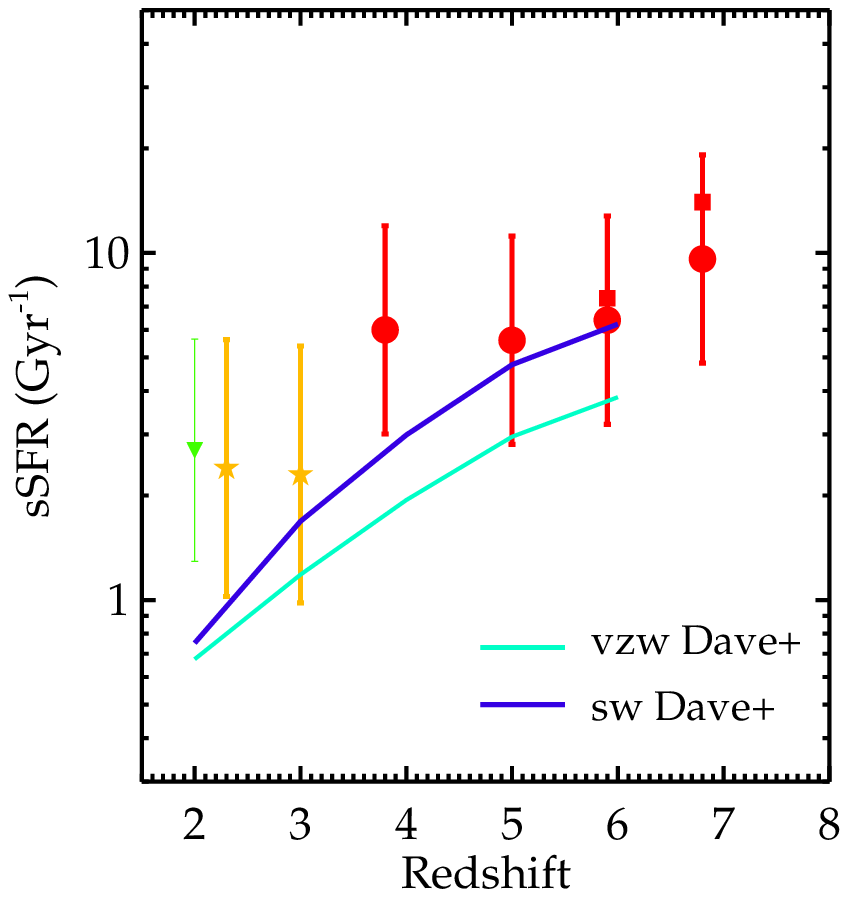}
\caption{{\it Left:} Evolution in the specific star formation rate (sSFR).  Our new measurements include  
stellar masses corrected for nebular emission line contamination. The solid red circles at $z>4$ show 
values derived assuming the nebular line EW distribution at $4<z<7$ remains identical to that derived 
at $3.8<z<5.0$ (Figure 5).  The solid squares correspond to values obtained when an evolving 
nebular line EW distribution is adopted.    The error bars show the assumed scatter about the mean sSFR 
taken from Reddy et al. (2012).   {\it Right:}   Comparison of observed sSFR to contemporary 
theoretical models.   The solid lines show the sSFR evolution predicted from cosmological simulations 
discussed in Dav\'{e} et al. (2011), with the blue line corresponding to their "slow wind" model and 
the light green line corresponding to the momentum driven wind "vzw" model.  These models provide 
adequate fits at the highest redshifts ($z>5$) but undershoot the observed values at $z\simeq 2-4$.}
\end{figure*}

The reduced stellar masses we infer in \S4 clearly will affect the evolution of the sSFR 
at $z>4$. To estimate the impact,  we compute the sSFR in fixed stellar mass bins using a similar 
approach as for the stellar mass function.    We draw a large number ($N\simeq 10^5$) of 
luminosities from the latest measures of the $z\simeq 4-7$ UV LFs (Bouwens et al. 2012a).   For each 
luminosity and redshift bin, we compute a stellar mass using the log M$_\star$-M$_{\rm{UV}}$ relationship 
derived in \S4.1.   We consider the case in which the strength of nebular emission is constant 
at $z\gsim 4$ and also that in which the emission line equivalent widths increase with redshift  
(e.g., Figure 6).  

The SFR is computed from M$_{\rm{UV}}$ through a series of steps.   We account for dust 
extinction using the UV continuum slopes.  For each realisation of the UV LF, 
we draw a UV slope, $\beta$, by adopting the redshift-dependent $\rm{\beta}-M_{\rm{UV}}$ 
scaling relationships (Bouwens et al. 2012b).    The UV slope is then converted 
to a dust attenuation factor at 1600~\AA~ via the Meurer et al. (1999) IRX-$\beta$ relation 
(A$_{1600}$=4.43+1.99$\beta$).   The UV luminosity is converted to 
SFR following the canonical Madau et al. (1998) and Kennicutt et al. (1998) relation L$_{\rm{UV}}$=
(SFR/M$_\odot$ yr$^{-1}$) 8.0$\times$10$^{27}$ ergs s$^{-1}$ Hz$^{-1}$.
This relationship assumes a 0.1-125 M$_\odot$ Salpeter IMF and constant star formation 
rate of $\gsim 100$ Myr.     Finally, by examining the SFR and M$_\star$ of these realizations, 
we compute the median sSFR of the four dropout samples with stellar mass of 5$\times$10$^{9}$ M$_\odot$.    Before 
examining the results of this calculation, we discuss two important issues that we have 
hitherto neglected.
 
First we consider how the sSFR is affected by scatter in the  log M$_\star$-M$_{\rm{UV}}$ relationship.    
Note that the sSFR will be {\it overestimated} if one merely uses the  log M$_\star$-M$_{\rm{UV}}$ 
relation without taking into account the abundant population of lower SFR objects with large 
M$_\star$/L$_{\rm{UV}}$ ratios.   This issue is dealt with in detail in Reddy et al.  (2012) for galaxies 
at $z\simeq 2-3$.   Unfortunately, as we discussed in \S5.1, the intrinsic scatter is very poorly constrained in UV-selected 
samples at $z\gsim 4$.   As a result, previous estimates of the sSFR at $z>4$ have 
not accounted for M$_\star$/L$_{\rm{UV}}$ scatter.   To estimate how this shortcoming would affect the 
sSFR, we add log M$_\star$-M$_{\rm{UV}}$ scatter to the LF realization method described above. 
If the 0.5 dex observed $z\simeq 4$ scatter reported in Gonzalez et al. (2011a) is entirely intrinsic, then the median sSFR 
would be reduced by 2.8$\times$ at $z\simeq 4$.  Note that one might find slightly different adjustments for 
the same scatter at other redshifts owing to evolution in the luminosity function.    In \S5.1, we suggested that the intrinsic scatter is likely lower 
as systematic uncertainties in modelling (including uncertainties in the nebular corrections) surely broaden the 
dispersion in the stellar masses at fixed UV luminosity.   In this case, fewer low SFR galaxies contribute to 
the sSFR distribution at fixed mass, resulting in a larger median sSFR.   For example, a scatter of 0.2 dex would 
translate into a reduction of just 1.2$\times$ with respect to the case of no scatter.  Physically, one may expect 
that the scatter at fixed luminosity  would increase somewhat between $z\simeq 7$ and $z\simeq 4$, as galaxies 
have had more time to undergo punctuated episodes of star formation which elevate both their star formation rates 
and mass off of the main sequence.   We consider these possibilities in our discussion below.

We now examine a second issue which pushes the sSFR in the opposite direction.  In particular, 
we examine how scatter (and perhaps systematic offsets) in the conversion between 
dust-corrected L$_{\rm{UV}}$ and SFR affect our sSFR determination.   The conversion between 
UV luminosity and SFR that we use is valid for galaxies with model ages in excess of 100 Myr.   
Above this age, the conversion factor changes little for an assumed constant star formation history.   But below 
100 Myr, a larger SFR is required to produce a fixed  L$_{\rm{UV}}$ (see Figure 25 of Reddy et al. 2012).   
For example, a galaxy with 
model age of 10 Myr  requires a 1.8$\times$ larger SFR to reproduce the same L$_{\rm{UV}}$ as a galaxy 
with 100 Myr.  With the reduced ages implied by the nebular corrections, it seems likely that such 
young systems are present in $z\gsim 4$ dropout samples.   Inclusion of dispersion in the model ages
will preferentially shift the median SFR to larger values, resulting in somewhat larger sSFR.   Furthermore, 
it is of course conceivable, if not likely, that such young systems will 
become more common both at higher redshift, requiring a systematic shift toward higher sSFR at earlier times. 
  
We now turn to the derived sSFR evolution, which is shown in Figure 9a.  First, ignoring the effect of 
nebular emission and the scatter discussed above, we find that the sSFR actually {\it decreases} 
marginally with redshift over 
$4<z<7$, similar to the findings of Bouwens et al. (2012b).   This is driven largely by the redshift-dependence 
of the UV continuum slope $\rm{\beta}~ versus~ M_{\rm{UV}}$ relationship.    Galaxies at higher redshifts 
have bluer UV continuua (e.g., Bouwens et al. 2012b, Finkelstein 
et al. 2012), reducing the dust-corrected SFR for a given M$_{\rm{UV}}$.   Considering the  
fixed M$_\star$/L$_{\rm{UV}}$ ratios assumed in previous studies 
(e.g., Stark et al. 2009, Gonzalez et al. 2010), it is straightforward to understand this result.   As we have 
discussed above, without nebular corrections, the data actually support a mild increase in the 
M$_\star$/L$_{\rm{UV}}$ ratios with increasing redshift at $z\gsim 4$ (Figure 7); if the M$_\star$/L$_{\rm{UV}}$ ratios 
weren't held fixed (and nebular emission not considered), one would have derived a more rapid 
decrease in sSFR at $z>4$.

Incorporating our corrections for nebular emission reduces the  M$_\star$/L$_{\rm{UV}}$ ratios in the 
$z\simeq 5-7$ LBG samples, increasing the sSFR in this redshift range.   If the nebular line 
EW distribution at $z\gsim 5$ is similar to that 
seen in Figure 5b,  we find that the sSFR of galaxies with fixed stellar mass begins to show evidence for  
positive evolution with redshift, with the $z\simeq 7$ value (9.6 Gyr$^{-1}$) 4$\times$ larger than that at 
$z\simeq 2$.  This can be viewed as a conservative estimate of the sSFR evolution.  As we have 
argued, however, it more likely that the equivalent width of H$\alpha$ and [OIII] {\it increase} in strength 
with redshift at $z\gsim 4$, consistent with the evolution seen at intermediate 
redshift (Fumagalli et al. 2012).  Under these assumptions,  the derived sSFR shows greater
 redshift evolution, with the $z\simeq 7$ sSFR (14 Gyr$^{-1}$) roughly 6 $\times$ greater than 
that at $z\simeq 2$ (Reddy et al. 2012).     While intrinsic scatter in the M$_\star$/L$_{\rm{UV}}$ 
ratios might bring these numbers down somewhat (perhaps explaining the excess seen at $z\simeq 4$), 
this is likely counteracted somewhat by scatter and/or systematic evolution in the SFR/L$_{\rm{UV}}$ ratios 
and possibly a shift toward reduced scatter in the M$_\star$/L$_{\rm{UV}}$ ratios at higher redshifts.

To summarize, with the new dust corrections (Bouwens et al. 2012b) and adjustments for nebular emission 
contamination, we now find evidence for a power law increase in the 
sSFR at $z\gsim 2$ that is much more consistent with theoretical expectations than previous 
observations indicated.   Both the absolute values and rate of increase of the sSFR we derive 
at $z\gsim 5$ are very similar to those predicted in the simulations of Dav\'{e} et al. (2011a).   
Intriguingly the sSFR at $2<z<4$ still remains moderately in excess of theoretical expectations.  
As we have discussed above, the $z\simeq 4$ estimate of the sSFR might come down somewhat 
owing to scatter in the M$_\star$/L$_{\rm{UV}}$.   But the $z\simeq 2-3$ sSFR measurements 
include the effects of scatter, so the discrepancy remains puzzling, especially in light of the 
emerging agreement at $z\gsim 5$.    Previous theoretical studies have focused on a variety of 
explanations for the tension at $z\simeq 2$, including a time-varying initial mass function 
(e.g., Dav{\'e} 2008, Narayanan \& Dav{\'e} 2012).   Continued efforts along these lines are 
required to simultaneously explain the high sSFR at $z\simeq 2$ along with the revised $z\gsim 4$ 
sSFR estimates (Fig. 9).  

Recall that previous indications of a nearly flat sSFR in fixed stellar mass bins at $z>2$ 
required a mechanism by which star formation is made increasingly inefficient 
at earlier times.    Possibilities included the inefficient formation of molecular hydrogen 
in metal-poor galaxies (e.g., Robertson\&Kravtsov 2008; Gnedin et al. 2009; Krumholz \& Dekel 2012), or 
an increase in the mass outflow rate per unit star formation with redshift.
The updated estimates of the $z>3$ sSFR  no longer obviously require a significant 
suppression of star formation in galaxies with stellar mass in excess of 10$^{9}$ M$_\odot$.
The current measurements seem consistent with a picture whereby the rapidly increasing 
baryon accretion rates translate into higher sSFR at earlier times.

As with the stellar mass density, there is room for considerable improvements to these estimates 
in the coming years.    By providing more individual detections of UV-faint galaxies, deeper 
{\it Spitzer} data will enable improved measurements of the slope and scatter of the log M$_\star$-M$_{\rm{UV}}$ 
relationship in the redshift range considered in Figure 9.    It is also of interest to extend these 
measurements to $z\simeq 8$.   In this redshift regime, [OIII] lies in the 4.5$\mu$m filter, while the 
3.6$\mu$m filter is devoid of strong lines.   Thus, with deep {\it Spitzer} data, the SEDs of $z\simeq 8$ 
systems enable a unique method of deciphering how the strength of nebular emission evolves 
over $5\lsim z\lsim 8$, one of the key uncertainties in the current analysis.    

\section{Summary and Conclusions}
\label{sec:conclusion}

Measurement of the evolving stellar mass and sSFR distributions have proven critical 
to our understanding of early galaxy assembly and the UV photon budget of reionization-era 
galaxies.   Recently, it has become clear that many of these early estimates might 
be significantly in error due to the contamination of the {\it Spitzer}/IRAC bandpasses by nebular emission 
lines.    Knowledge of the strength of these emission lines is necessary for robust determinations 
of the stellar mass density and sSFR evolution.  As these emission lines are shifted out of the observed 
atmospheric window, direct spectroscopic measurements will not be feasible until JWST.  

In this paper, we present a method which enables constraints on nebular emission at $z>4$ by 
combining large spectroscopic samples and deep {\it Spitzer} photometry.   Like Shim et al. (2011), 
we focus on the redshift range $3.8<z<5.0$, over which  the IRAC 
[3.6] filter is contaminated by strong emission lines (H$\alpha$, [NII], [SII]) while the [4.5] filter is 
free of nebular contamination.  Examining a carefully-selected subset of 45 galaxies, we find that the 
3.6$\mu$m flux is systematically in excess of the expected stellar continuum flux, revealing the presence of 
strong nebular emission.  No excess is seen in a spectroscopic sample at $3.1<z<3.6$, a redshift range 
over which no strong emission lines contaminate the IRAC filters.   We use the photometric excesses 
in the contaminated [3.6] filter to estimate the equivalent width distribution of H$\alpha$ emission 
at $3.8<z<5.0$.   Equipped with this measure of nebular emission at high-redshift, we re-evaluate 
the evolution in the sSFR and stellar mass density at $z\gsim 4$.   Our primary conclusions from this 
analysis are summarized below.
 
1.  We find that the mean equivalent width of emission lines contaminating the [3.6] filter 
is $\rm{<log_{10} (W_{3.6}/\AA)>}$ $\simeq 2.57-2.73$.   
We estimate that $\simeq 76$\% of this signal arises from H$\alpha$, implying an average H$\alpha$ 
equivalent width of $\rm{<log_{10} (W_{H\alpha}/\AA)>}$ 
$\simeq 2.45-2.61$ at $3.8<z<5.0$.  
 
2.  The mean H$\alpha$ equivalent width inferred at $3.8<z<5.0$ appears greater than that  
for similar star-forming samples at lower redshifts.   While definitive knowledge of the evolution in the 
H$\alpha$ EW surely awaits direct spectroscopic measurement,   the evolution we infer over 
$2\lsim z\lsim 5$ is certainly consistent 
with the (1+z)$^{1.8}$ power law derived in Fumagalli et al. (2012).   This likely reflects 
an increase in the sSFR at $z\gsim 2$, and importantly suggests that the EW of 
nebular emission continues to increase at $z\gsim 5$.
 
3.  Using the H$\alpha$ EW distribution we derive at $3.8<z<5.0$, we explore 
how nebular contamination is likely to affect the physical properties of galaxies at $z>3$.  
We find that the stellar masses are reduced, on average, by 1.1, 1.3, 1.6, and 2.4$\times$ for 
dropout samples with mean redshifts of $z\simeq \rm{4,~5,~6,~and~7}$, respectively.   If the 
equivalent widths of nebular lines continue to increase in amplitude at $z\gsim 5$, we estimate 
that the reduction in the stellar masses are likely to increase to 1.9 and 4.4$\times$ at $z\simeq 6$ and 7.  
We note that these corrections are representative for average measures of the dropout populations and not individual 
galaxies.

4.   As the stellar mass density provides a valuable integrated measure of early star formation, 
constraints on the level of nebular contamination are critical to our knowledge of the ionizing 
photon budget of galaxies throughout the reionization era.    After correcting for nebular emission 
contamination, we find a factor of $\simeq 2\times$ reduction from previous estimates.  The downward  
revisions to the stellar mass density improve consistency with expectations from the integrated star 
formation rate density.     Extending such nebular-corrected measurements to emerging galaxy 
samples at $z\simeq 8-9$ will yield an integral constraint on the UV photon budget 
during $z\simeq 10-15$.

5.    Whereas previous derivations showed little evolution in the sSFR of fixed mass galaxies over 
$2<z<7$,  we demonstrate that after accounting for nebular emission and correcting for dust, 
the sSFR increases by 4-6$\times$ over $2<z<7$.    The absolute sSFR values inferred at 
$z\gsim 5$ appear largely similar with predictions from simulations.   While there certainly 
remains room for improvement (in both the data and the modelling), the increase in the sSFR at 
$z\gsim 4$ seems consistent with a picture whereby increasing baryon accretion rates at larger 
redshift translate into larger sSFR in galaxies of a fixed stellar mass. 
 
\subsection*{ACKNOWLEDGMENTS}

We are grateful to Rychard Bouwens, Romeel Dav\'{e}, Desika Narayanan, Ivo Labb\'{e}, Masami 
Ouchi, Naveen Reddy, Daniel Schaerer, and Valentino Gonzalez for useful conversations.  
We thank Romeel Dav\'{e} for making the results of his simulations available to us.  
DPS acknowledges support from NASA through Hubble Fellowship grant \#HST-HF-51299.01 awarded 
by the Space Telescope Science Institute, which is operated by the Association of Universities for 
Research in Astronomy, Inc, for NASA under contract NAS5-265555.   BER is partially supported through STScI grant HST-GO-12498.12-A. Support for Program number HST-GO-12498.12-A was provided by NASA through a grant from the Space Telescope Science Institute, which is operated by the Association of Universities for Research in Astronomy, Incorporated, under NASA contract NAS5-26555.   JSD acknowledges the support of the European Research Council via the
   award of an Advanced Grant. JSD and RJM acknowledge the support of the
   Royal Society via a Wolfson Research Merit Award and a University Research
   Fellowship respectively.    Some of the data presented herein were obtained at the W.M. Keck Observatory, which is operated as a scientific partnership among the California Institute of Technology, the University of California and the National Aeronautics and Space Administration. The Observatory was made possible by the generous financial support of the W.M. Keck Foundation.

\clearpage

\begin{deluxetable}{llllll}
 \tabletypesize{\scriptsize}
\tablecolumns{6}
\tablecaption{GOODS bright $3.8<z<5.0$ spectroscopic sample}
\tablehead{
\colhead{ID} &
\colhead{RA(J2000)} &
\colhead{DEC(J2000)} &
\colhead{z$_{\rm{spec}}$} &
\colhead{z$_{\rm{850}}$} &
\colhead{$\Delta  [3.6]$}  
}
\startdata
S44\_1649 &      03:32:05.022 &      -27:46:12.65 &        3.91  &       24.47  &       0.07 \\
S43\_2212 &      03:32:06.615 &      -27:47:47.69 &        3.94  &       24.28  &       0.13 \\
S33\_6294 &      03:32:14.497 &      -27:49:32.69 &        4.74  &       25.40  &       0.26 \\
S33\_8715 &      03:32:18.257 &      -27:48:02.53 &        4.28  &       24.65  &       0.45 \\
S33\_15763 &     03:32:27.939 &      -27:46:18.57 &        4.00  &       25.23  &       0.17 \\
S23\_20730 &     03:32:34.349 &      -27:48:55.81 &        4.14  &       24.11  &       0.22 \\
S24\_23979 &     03:32:38.729 &      -27:44:13.34 &        4.00 &       24.81  &        0.00  \\
S23\_24940 &     03:32:40.086 &      -27:49:01.21 &        4.13 &       26.45 &         0.35 \\
S24\_25118 &     03:32:40.385 &      -27:44:31.00 &        4.13 &       25.24 &         0.04 \\
S22\_25614 &     03:32:41.159 &      -27:51:01.50 &        4.06 &       25.25 &         0.42 \\
S23\_28451 &     03:32:46.247 &      -27:48:46.99 &        4.02 &       24.88  &        0.39 \\
S12\_29436 &     03:32:48.244 &      -27:51:36.90 &        4.36 &       24.87 &         0.67  \\
S13\_31908 &     03:32:54.035 &      -27:50:00.81 &        4.43 &       25.07 &        -0.25 \\
S12\_32366 &     03:32:55.249 &      -27:50:22.46  &        4.17 &       24.42  &       0.20  \\
S12\_33166 &     03:32:58.380 &      -27:53:39.58  &        4.40 &       25.75 &        0.14  \\
S43\_1669 &      03:32:05.080 &      -27:46:56.52  &        4.82 &       23.79 &        0.14 \\
S44\_1745 &      03:32:05.259 &      -27:43:00.42  &        4.80 &       25.24 &        0.14 \\
S45\_3792 &      03:32:10.027 &      -27:41:32.65  &        4.81 &       25.03 &        0.27 \\
S34\_11180 &     03:32:21.931 &      -27:45:33.07  &        4.79 &       25.82 &        0.17 \\
S33\_11861 &     03:32:22.884 &      -27:47:27.57  &        4.44 &       24.93 &        0.43 \\
S33\_11915 &     03:32:22.971 &      -27:46:29.08  &        4.50 &       25.34 &        0.33 \\
S35\_16226 &     03:32:28.563 &      -27:40:55.74  &        4.60 &       25.44 &        0.24 \\
S31\_16819 &     03:32:29.291 &      -27:56:19.46  &        4.76 &       25.05 &        0.10 \\
S22\_20041 &     03:32:33.475 &      -27:50:30.00  &        4.90 &       25.77 &        0.25 \\
S24\_24961 &     03:32:40.118 &      -27:45:35.47  &        4.77 &       25.55 &        0.45  \\
S21\_26522 &     03:32:42.623 &      -27:54:28.95  &        4.40 &       25.61 &        0.58 \\
S12\_32900 &     03:32:57.169 &      -27:51:45.01  &        4.76 &       24.64 &        0.40 \\
N33\_14884 &     12:36:42.235  &      +62:15:22.93   &       4.42  &       24.47   &        0.27 \\
N14\_27206 &     12:37:57.510  &      +62:17:18.77   &       4.71  &       23.82   &        0.28 \\
N42\_5352 &      12:36:14.513  &      +62:11:40.61   &       4.15  &       25.31   &        0.30 \\
N42\_12760 &     12:36:36.823  &      +62:12:04.03  &        3.90   &       24.94  &        0.35 \\
N33\_20202 &     12:36:55.940  &      +62:14:12.44  &        3.91   &       23.78  &        0.15 \\
N33\_25472 &     12:37:09.840  &      +62:14:39.37  &        4.25   &       25.01  &        0.32 \\
N23\_28987 &     12:37:19.688  &      +62:15:42.46  &        4.53   &       25.49  &        0.27 \\
N34\_21578 &     12:36:59.377  &      +62:19:05.41  &        3.86   &       25.30  &        0.10 \\
N34\_21756 &     12:36:59.758  &      +62:18:54.33  &        3.86   &       24.78  &        0.27 \\
N34\_23754 &     12:37:05.013  &      +62:17:31.01  &        3.93   &       24.62  &        0.31 \\
N35\_26133 &     12:37:11.814  &      +62:22:12.30  &        4.05   &       24.25  &        0.27 \\
N35\_26600 &     12:37:13.037  &      +62:21:11.16  &        4.05   &       24.09  &        0.33 \\
N24\_28740 &     12:37:19.003  &      +62:19:53.51  &        4.19   &       24.38  &        0.18 \\
N25\_29248 &     12:37:20.446  &      +62:22:14.85  &        4.05   &       24.63  &        0.58 \\
N24\_29391 &     12:37:20.845  &      +62:18:43.22  &        4.07   &       25.42  &        0.41  \\
N32\_22884 &     12:37:02.520  &      +62:11:55.00  &        4.02   &       25.64  &        0.10  \\
N24\_27374 &     12:37:15.103  &      +62:20:05.21  &        4.06   &       25.85  &        0.35 \\
N42\_8958 &      12:36:25.972  &      +62:08:59.43  &        4.14   &       24.36  &       0.49 \\
\enddata
\end{deluxetable}

\end{document}